\documentclass[twocolumn,aps,pra]{revtex4-1}

\usepackage{amsmath}
\usepackage{float}
\usepackage{graphicx}
\usepackage[breaklinks]{hyperref}
\usepackage{breakurl}
\usepackage{multirow}
\usepackage{natbib}
\usepackage{subcaption}

\newcommand{\Project}{{https://github.com/sabauma/molecular-dynamics.git}}
\newcommand{\degree}{\ensuremath{^\circ}}

\begin{document}

\author{Spenser Bauman}
\email{sabauma@gmail.com}
\affiliation{Department of Computer Science, Penn State University}

\author{Max Fomitchev-Zamilov}
\email{max@quantum-potential.com}
\affiliation{Department of Computer Science, Penn State University}

\title{Effects of Gas Dynamics on Rapidly Collapsing Bubbles}
\date{\today}

\begin{abstract}
    The dynamics of rapidly collapsing bubbles are of great interest due to
    the high degree of energy focusing that occurs within the bubble.
    Molecular dynamics provides a way to model the interior of the bubble and
    couple the gas dynamics with the equations governing the bubble wall.
    While much theoretical work has been done to understand how a bubble
    will respond to an external force, the internal dynamics of the gas system
    are usually simplified greatly in such treatments.
    This paper shows how the gas system dynamics affect bubble collapse and
    illustrates what effects various modeling assumptions can have on the
    motion of the bubble wall.
    In addition, we present a method of adaptively partitioning space to
    improve the performance of collision intersection calculations when using
    an energy dependent collision cross section.
\end{abstract}

\maketitle

\section{Introduction}

The motion of a collapsing bubble in a sound field has been studied
extensively using hydrodynamics models \cite{yuan_physical_1998,
lofstedt_toward_1993, keller_damping_1956, gaitan_sonoluminescence_1992,
prosperetti_bubble_1986, lezzi_bubble_1986}.
Experimental results have shown that a purely hydrodynamical model can often
provide a very accurate description of bubble motion during most of the bubble's
collapse, but provide less insight into the internal mechanics of the bubble
during the latter stages of collapse.

More recently, molecular dynamics models have been applied to understand the
behavior of gas trapped in a collapsing bubble \cite{metten_molecular_2000,
bass_molecular_2008, ruuth_molecular_2002, BassPhd, woo_molecular_1999,
gaspard_imploding_2004}.
Molecular dynamics models allow for a more accurate description of the
internal conditions of the bubble at the cost of greater computational
resources.

The motion of a gas bubble in a liquid is described by a Rayleigh-Plesset
style equation \cite{putterman_sonoluminescence_2000}: a second order,
non-linear differential equation describing the motion of the bubble
wall in response to external driving forces as well as the force exerted by
the gas on the bubble wall.
In the context of the Rayleigh-Plesset equation, the gas is typically modeled
using various equations of state.
Due to the high compression involved, van der Waals (VDW) equations of state
are often used, but more realistic models of gas dynamics have also been
studied \cite{yuan_physical_1998}.
However, accurately modeling the gas dynamics analytically is difficult due to
the extreme conditions experienced within the bubble.

Previous molecular dynamics models of bubble cavitation have made assumptions
to ease modeling of the bubble wall, such as constant velocity wall motion
\citep{gaspard_imploding_2004}, or model the gas system while separately
modeling the bubble wall using idealized equations of state
\citep{ruuth_molecular_2002, bass_molecular_2008, BassPhd}.
The work of \cite{kim_validation_2008} made use of a molecular dynamics
simulation coupled with the bubble wall equation to model the collapse of the
bubble, but the model used scaled atomic parameters and modeled particles as
constant diameter hard spheres.

In the following work, we will describe the model of both the gas dynamics and
the bubble wall and describe how they are coupled.
We will show how various modeling choices and assumptions affect the evolution
of the system as a whole (i.e.\ the evolution of the bubble as well as the
state of the gas particle system), in the hopes of providing a more accurate
description of bubble collapse.
Due to the coupling of bubble and gas dynamics, the choice of microscopic gas
properties now has an effect on the macroscopic property of bubble evolution,
providing a potential avenue for validation of various modeling assumptions
and choices.

The aim of this work is to understand how the motion of the bubble wall is
influenced by modeling choices and physical properties of the gas system being
simulated, and how this feed back can influence the gas inside of the bubble.

\section{Modeling}

The model presented here based on the works of \cite{ruuth_molecular_2002,
BassPhd}, where an event-driven molecular dynamics simulation was employed to
study the behavior of the gaseous interior of a collapsing bubble.
This section will serve as an overview of the model and how it differs from
previous work.

\subsection{Sphere Properties}
\label{subsec:sphere_mechanics}

Within the bubble, all gas species are modeled as spheres undergoing impulse
only during collisions with each other or with the bubble wall; all other
motion is purely inertial.
Conveniently, this allows for the precise integration of the system of
particles, as all relevant events in the simulation can be computed precisely.
In particular, particle-particle collisions can be accurately determined.

To achieve a more accurate model in the high energy state that the bubble
attains, we use the Variable Soft Sphere (VSS) \citep{matsumoto_variable_2002,
koura_variable_1992} model to determine the collision cross sections of
particle pairs in the simulation.
The energy dependent diameter of the VSS model is given by
\[
    \sigma =
    \left(
    \frac{5(\alpha + 1)(\alpha + 2) (m / \pi)^{1/2} (k T_{ref})^{\omega + 1/2}}
         {16 \alpha \Gamma(4 - \omega) \mu_{ref} E_t^{\omega}}
    \right)^{1/2},
\]
with $k$ being Boltzmann's constant, $m$ the mass of the particle, $\omega$
the viscosity index, and $\alpha$ a dimensionless constant specific to each gas.
Additionally, $E_t = (1/2) m_r c_r^2$ is the asymptotic kinetic energy, $m_r$
is the reduced mass, and $c_r$ is the relative velocity for the particle pair.

\begin{table}
    \scriptsize
    \centering
    \begin{tabular}{lcccr}
        \hline \hline
        \multicolumn{5}{c}{Exact Values} \\ \hline
        Gas   & Mass (g/mol) & $\omega$ & $\mu_{ref} (N \cdot s \cdot m^{-2})$
            & $\alpha$ \\ \hline

        D$_2$ & $4.03$       & $0.199$  & $1.185$ & $1.420$ \\
        He    & $4.00$       & $0.202$  & $1.865$ & $1.431$ \\
        Ar    & $39.95$      & $0.201$  & $2.117$ & $1.425$ \\
        Xe    & $131.29$     & $0.302$  & $2.107$ & $1.165$ \\
        \hline
        \multicolumn{5}{c}{Adopted Values} \\ \hline
        Gas & Mass (g/mol)   & $\omega$ & $\mu_{ref} (N \cdot s \cdot m^{-2})$
            & $\alpha$ \\ \hline
        D   & $2.01$         & $0.199$  & $1.185$ & $1.420$ \\
        \hline \hline
    \end{tabular}
    \caption{Variable soft sphere parameters for various gases.
        Note the division between known values and those that have been
        adopted from their diatomic counterparts.}
    \label{tab:gas_parameters}
\end{table}

Several of the parameters (see Table \ref{tab:gas_parameters}) for the VSS
model ($\alpha$, $\mu_{ref}$, and $\omega$) were derived empirically.
Unfortunately, the appropriate constants are not available for deuterium, in
either its diatomic or monatomic form.
The constants for the diatomic gas species are used for the monatomic
counterpart as well, since those values are not available.
In effect, the only properties of the particle that are altered by
dissociation are the mass, ionization energy, and collision cross section (due
to the cross section's dependence on particle mass).
The effects of this are fairly straightforward to analyze.
Given two particle of the same species colliding, the effect of halving the
mass is to decrease the collision cross section by a factor of
$\sqrt[4]{0.5} \approx 0.841$, assuming the asymptotic kinetic energy remains
constant.
This is the best that can be done, in the absence of the appropriate VSS
parameters for the monatomic species.

Due to the method used to determine pressure, the use of a more realistic
collision cross section becomes more important; the communication between the
bubble wall and the gas makes an accurate gas model increasingly important
(outlined in Section \ref{subsec:gas_pressure}).

From the collision cross section described above, the time at which two
particles collide can be computed as the solution of
\[
    \left| \mathbf{r} + \mathbf{v} t \right| = \sigma,
\]
where $\mathbf{r}$ and $\mathbf{v}$ are the relative position and velocity
vectors of the two particles, respectively.
Collisions are then carried out so as to preserve the energy and momentum of
the system.
The change in velocity is given by
\[
    \Delta \mathbf{v}_1 = - \Delta \mathbf{v}_2 =
    -2 \mu \, \textrm{proj}_{\mathbf{r}} \mathbf{v},
\]
at the time of the collision.
$\Delta \mathbf{v}_1$ and $\Delta \mathbf{v}_2$ are the change in velocity of
the first and second particles, respectively, and $\mu$ is the reduced mass of
the two particles.

\subsection{Wall Conditions}
\label{subsec:wall_conditions}

Heat exchange between the gas and the surrounding liquid is modeled using
a heat bath boundary condition described by \citet{BassPhd}, rather than that
used by \citet{kim_validation_2008}.
The surrounding liquid is assumed to maintain a constant temperature and the
outgoing velocity vector of the particle is chosen to correspond to the
velocity distribution of the gas having the same temperature as the bubble
wall ($T_0$).

Both the tangential and normal components of the resulting velocity vector
come from the two dimensional Maxwell-Boltzmann distribution with the
probability density function
\[
    f(v) = v e^{- \frac{mv^2}{2kT_0}}.
\]

The resulting vector is given as the sum of two such vectors plus the wall
velocity and a random azimuth angle chosen uniformly on the interval
$[0, 2 \pi)$.
In contrast, the wall used by \citet{kim_validation_2008} chose the outgoing
vector based on specular reflection but with a magnitude corresponding to the
temperature of the wall.

The effects of altering the direction of outgoing particles has not been
explored, but it has been noted by \citet{ruuth_molecular_2002} that the
choice of angular distribution can influence how particles build up along the
wall.
While no qualitative differences were found in that particular work, the
buildup of particles along the bubble wall has a definitive effect when
accounting for pressure due to the gas.

The thermal wall conditions simulate heat exchange between the gas and the
liquid.
At various times in the simulation, the thermal conditions can have either
a heating or cooling effect on the gas contained in the bubble.
During the expansion phase of the bubble, the thermal wall conditions keep the
internal temperature of the gas near that of the wall, preventing it from
cooling as drastically as it would under purely adiabatic expansion.
The wall has a cooling effect during the collapse of the bubble.
When the collapse velocity is low, the gas and the liquid remain in a more or
less equilibrium state.
At higher velocities, the gas starts heating faster than the bubble wall can
conduct heat away from the gas, and the formation of the shock wave allows the
central gas to heat up without significant heat loss to the surrounding
liquid.

For comparison, a second wall method will be included, where particles
incident on the bubble wall undergo a specular reflection which conserves
energy.
These two methods exist at opposite ends of the spectrum when considering the
energy a particle loses to the bubble wall.
While the specular wall results in the particle having the same amount of
energy before and after the impact, the thermal wall assumes that the particle
will always thermalize with the bubble wall.
As noted by \citet{ruuth_molecular_2002} and later explored by
\citet{kim_validation_2008}, the reality will have some properties of both of
these models.

\subsection{Bubble Collapse}
\label{subsec:bubble_collapse}

The bubble wall is modeled using a modified version of the Keller-Miksis
\citep{gaitan_sonoluminescence_1992, yuan_physical_1998} formulation of the
Rayleigh-Plesset (denoted KM from here on) equation which factors in shock
formation due to liquid compressibility, viscous damping, and surface tension.
Previous efforts have made use of either the unmodified Rayleigh-Plesset
equation \cite{ruuth_molecular_2002} or the Keller-Kolodner formulation of the
Rayleigh-Plesset equation \cite{BassPhd, metten_molecular_2000}.
The Keller-Kolodner formulation includes acoustic damping effects, but is only
a zeroth order approximation in the Mach number of the bubble wall
\citep{prosperetti_bubble_1986}.
While liquid compressibility only becomes significant for a short period of
time during the overall collapse, it can have a significant effect on the
formation of shock waves.
In particular, poorly modeling liquid compressibility can result in
significantly higher maximum bubble-wall velocities and internal temperatures
\cite{yuan_physical_1998}.
Work by \citet{lezzi_bubble_1986} shows that this and similar first-order
approximations will tend to slightly overestimate the energy lost to acoustic
damping, but the effect becomes more important at higher Mach numbers.

The Keller-Miksis equation used to model the bubble driven by either
a constant or sinusoidal pressure function which is spatially homogeneous.
\begin{align*}
    &\left(1 - M \right) \rho R \ddot{R}
    + \frac{3}{2} \left(1 - \frac{M}{3} \right) \rho \dot{R}^2 = \\
    &\left( 1 + M \right) \left[P_b(R,t) - P_\infty - P_a \left(t + t_R \right)
    \right] + t_R \frac{\partial P_b(R,t)}{\partial t},
\end{align*}
where $P_b(R,t) = P_g(R,t) - 2 \sigma / R - 4 \mu (\dot{R} / R)$ and
$t_R = \frac{R}{c}$.
Here, $P_g$ is the pressure exerted by the gas on the bubble wall, $P_0$ the
ambient liquid pressure, $P_d$ the driving pressure, $\sigma$ is the surface
tension of the liquid, $\rho$ is the density of the liquid, $\eta$ is the
dynamic viscosity of the liquid, and $M$ is the Mach number of the bubble
wall; $R$ is the radius of the bubble and over dots denote derivatives with
respect to time.
The maximum collapse velocity under the KM equation is significantly less than
that of the unmodified RP equation or the Keller-Kolodner formulation, due to
the damping effects of liquid compressibility.

The KM equation is then solved numerically by a variable time step ODE solver
using a Runge-Kutta Dormand-Prince method.
The time step of the solver is used to discretize the simulation into
\emph{snapshots}, which comprises all of the events between two steps of the
KM equation taken by the ODE solver.
This works surprisingly well, because the ODE solver scales its step size to
keep the error within the proscribed bounds.
This naturally results in smaller time steps when things are happening more
quickly within the simulation.

\subsection{Gas Pressure}
\label{subsec:gas_pressure}

Coupling the molecular dynamics component of the model with the bubble wall
was discussed in \cite{ruuth_molecular_2002}, and code to empirically
determine the gas pressure at the bubble wall was implemented in
\cite{BassPhd} but was not tied into the equation governing the motion of the
bubble wall.
The method used by Bass involved a linear interpolation of two time regions
averaged over a large number of particle-wall interactions.
Our own efforts have found that methods involving a simple linear
interpolation of the pressure do not work well when coupled with the bubble
wall equation, as they do not accurately capture the asymptotic rise in
pressure when attempting to extrapolate the pressure for times past which we
have data for; this results in an unphysical level of collapse due to pressure
underestimation.
Here, we present an improved method to empirically determine the gas pressure
at the bubble wall and show it to deviate from adiabatic compression only at
high velocities.

Previous efforts \citep{metten_molecular_2000, bass_molecular_2008,
ruuth_molecular_2002, BassPhd} at an MD simulation of bubble cavitation have
have coupled various forms of the Rayleigh-Plesset equation with the VDW
adiabatic equation of state
\[
    P_g(R) = \frac{P_0 R_0^{3 \gamma}}{(R^3 - a^3)^\gamma}
\]
where $\gamma = 5/3$ for a monoatomic gas, and $a$ is the VDW hard core radius
for the bubble.
Such a treatment ignores the fact that the gas inside of the bubble is
non-homogeneous in temperature and density for much of the collapse owing to
the formation of strong shocks within the bubble.

Fortunately, the simulation has direct access to all interactions between the
bubble wall and the gas contained therein.
For each particle that collides with the wall, we record the change in
momentum along the normal vector per unit of area
\[
    \Delta \rho = \frac{(\vec{v}_{out} - \vec{v}_{in}) \cdot \vec{n}}{A},
\]
along with the time of each particle-wall interaction.

These events are recorded for the duration of one snapshot and then used to
estimate the pressure for the next snapshot.
To do so, the total change in momentum per unit of time is computed as the sum
of all such $\Delta \rho$ over the duration between the $n$ recorded wall
impacts.
This computed value $P_{\bar{t}}$ is taken to be the pressure at the mean time
of all the particle-wall interactions $\bar{t}$,
\[
    P_{\bar{t}} = \frac{\sum_{i=1}^n \Delta \rho}{t_n - t_1}.
\]
$P_{\bar{t}}$ then, gives is the pressure of the bubble at time $\bar{t}$
corrected for heterogeneities in the simulated gas.
Within a given snapshot, we assume that pressure grows according to adiabatic
compression as
\[
    P_g(R) = P_{\bar{t}} \left( \frac{R_{\bar{t}}}{R} \right)^{3 \gamma},
\]
where $R_{\bar{t}}$ is the radius of the bubble at time $\bar{t}$.

Therefore, this method models the gas pressure as a corrected adiabatic
pressure function, where the interaction between the bubble wall and the gas
is used to factor in the heterogeneities of the gas during collapse.

During the initial phase of collapse, when the wall velocity is relatively
low, the discrepancy in the KM equation between the VDW pressure and the
empirically determined pressure is negligible.
The deviation occurs only in the final moments of collapse, when high collapse
velocity of the bubble wall induces the formation of shock waves and mass
segregation.

As will be shown in Section \ref{sec:results}, the pressure gauge results in
a marked change in the overall RP curve during the final moments of collapse.
With the inclusion of the pressure gauge, the importance of the choice of
certain gas properties increases significantly, as the properties of the gas
alter its response to the bubble wall, which in turn affects the pressure
exerted on the wall.

\subsection{Ionization and Dissociation}
\label{subsec:ionization_and_dissociation}

As the energy inside of the bubble increases, the interior reaches
temperatures sufficient to ionize the gas particles and to dissociate any
diatomic gases contained therein.
To facilitate ionization of the gases, each particle's ionization level is
tracked.
When a collision occurs, the energy at the center of mass frame is computed;
if that energy is sufficient to ionize one of the gas particles, that energy
is subtracted from the pair of particles when computing the velocity vectors
from the collision.
Since we do not track the resulting electrons, none of the remaining energy is
alloted to the electron.

\begin{table}
    \scriptsize
    \centering
    \begin{tabular}{lccccccccc}
        \hline \hline
        Gas    & \multicolumn{8}{c}{Ion} \\ \cline{2-9}
            & Neutral & $+1$ & $+2$ & $+3$ & $+4$ & $+5$ & $+6$ & $+7$ \\ \hline
        D                                     & 1.31    &      &      &
            &      &      &      & \\
        D$_2$ \cite{watanabe_comparison_2010} & 1.49    &      &      &      &
            &      &      & \\
        He \cite{ruuth_molecular_2002}        & 2.37    & 5.25 &      &      &
            &      &      & \\
        Ar \cite{ruuth_molecular_2002}        & 1.52    & 2.67 & 3.93 & 5.77
            & 7.24 & 8.78 & 12.0 & 13.8 \\
        Xe \cite{ruuth_molecular_2002}        & 1.17    & 2.05 & 3.10 & 4.60
            & 5.76 & 6.93 & 9.46 & 10.8 \\
        \hline \hline
    \end{tabular}
    \caption{Ionization potential in (MJ/mol). The entries indicate the amount
        of energy required to strip an electron from a given state.
        The values for D are assumed to be the same as for H, which can be
        computed directly using Rydberg's equation.}
    \label{tab:ionization_potential}
\end{table}

To more accurately model diatomic gases such as H$_2$ and D$_2$, we also
consider the possibility that a collision will result in the dissociation of
any diatomic gas particles involved in the collision.
The method for recognizing the occurrence of dissociation is the same as for
ionization.
If the energy at the center of mass frame is sufficient to cause dissociation,
that energy is subtracted from the system and then the outgoing vectors for
the two particles are computed.
The process differs once the outgoing vectors are computed and assigned to the
new particles.
First, the particle that dissociated is demoted to its monatomic equivalent.
Next, a new particle is introduced into the simulation translated an atomic
diameter towards the bubble center relative to the original particle.
Translating the particle prevents them from overlapping and the direction is
chosen to alleviate the need to determine whether the new particle is being
placed outside of the bubble.
The two resulting particles have the same velocity vector as the original,
diatomic particle, which ensures that energy and momentum are conserved.
Currently, deuterium is the only diatomic gas used in the simulations and is
assumed to have the same dissociation energy as diatomic hydrogen ($0.436$
MJ/mol).

The introduction of ionization and dissociation has the effect of cooling the
gas during collapse.
Dissociation can also have the effect of enhancing mass segregation, as
segregation is primarily a result of a mass disparity between the heavy and
light particles \citep{bass_molecular_2008}.
Unless otherwise stated, all simulations are include the effects of ionization
and dissociation where applicable.

\subsection{Initialization}
\label{subsec:initialization}

Before the simulation can begin, the bubble must be populated with gas
particles representative of the conditions expected at time zero if the bubble
were given time to equilibrate completely with its surroundings.
Therefore, the bubble is initialized to the appropriate, uniform density with
the velocity distribution of the particles chosen corresponding to the ambient
temperature of the liquid.

The density of the bubble is determined by the ambient radius of the bubble,
which is to say the radius of the bubble which would result in no expansion of
compression of the bubble with the given ambient pressure and temperature if
no driver is applied.
This condition is given by the ideal gas equation as
\[
    n = \frac{\left(P_\infty + \frac{\sigma}{R_a} \right) V_a}{R T_\infty},
\]
where $n$ is the number of particle in the system, $P_\infty$ is the ambient
pressure of the liquid, $R_a$ is the radius of the bubble, $V_a$ is the
ambient volume of the bubble, $R$ is the ideal gas constant, and $T_\infty$ is
the ambient temperature of the gas.
This determines the number of particles which inhabit a bubble of the given
ambient conditions.
The particles are initially placed in a three dimensional cubic lattice whose
size is determined by the initial size of the bubble, which does not have to
be the ambient size of the bubble.

The velocity vectors for each particle are assigned randomly based on the
Maxwell-Boltzmann distribution.
Each of the three axes are assigned a velocity from a normal distribution with
a mean of zero and a variance of $\sigma^2 = k T_\infty / m$, where $m$ is the
mass of the particle in question and $k$ is Boltzmann's constant.

\section{Algorithm}
\label{sec:algorithm}

The algorithm presented below consists of a modified version of that found in
\cite{bass_molecular_2008,ruuth_molecular_2002}.
The same efficient, event driven algorithm with spatial partitioning as
\citep{rapaport_art_2004} and supplemented with conical symmetry reduction
\citep{bass_symmetry_2008} to integrate the particle system.
All simulation results presented here use a cone angle of 15 degrees, which
was chosen based on \citet{bass_symmetry_2008} to give dependable results
while drastically reducing the number of particles being simulated.
Using these methods and making some improvements, a system of a million
particles can be fully collapsed within 24 hours, though this depends heavily
on the composition and conditions it is subjected to.

Benchmarking was performed using an Alienware laptop with an Intel i7-3610QM
processor running at a base speed of 2.30GHz (max of 3.3GHz) supporting up to
8 hardware threads and 12GB of RAM.
Bulk processing was facilitated by an AMD workstation with dual AMD Opteron
6284SE processors operating at a base speed of 2.70 GHz (max of 3.4GHz) with
a total of 32 CPU cores and 32GB of RAM.

The project was implemented in C++ making use of the Boost project.
The code can be found on \href{https://github.com}{GitHub} at
\href{\Project}{\Project} and is freely available for use and review.

\subsection{Spatial Partitioning}
\label{subsec:spatial_partitioning}

\begin{figure*}[t]
    \centering
    \begin{subfigure}[b]{0.48\textwidth}
        \centering
        \includegraphics[width=\textwidth]{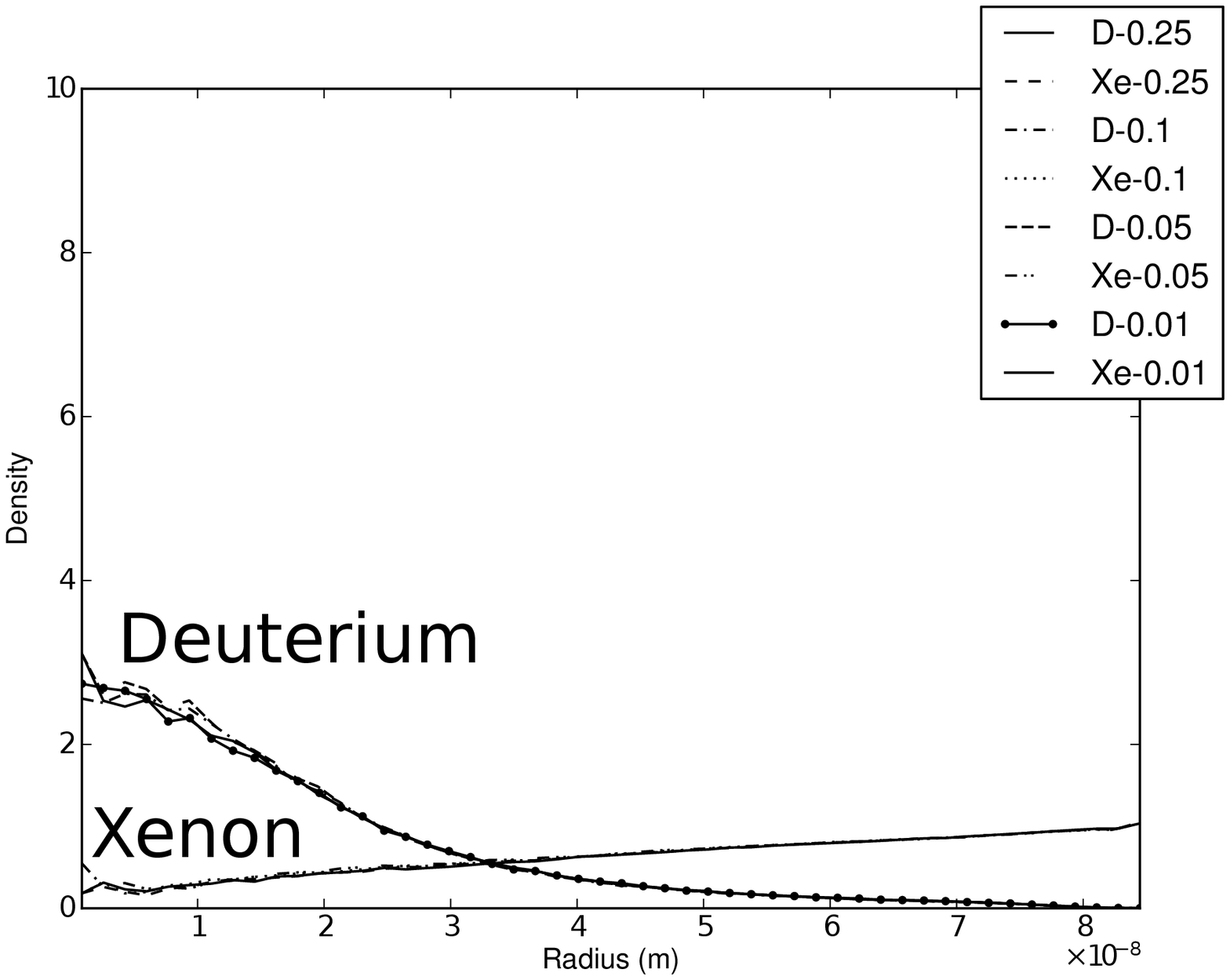}
        \caption{Density vs Radius}
        \label{fig:error_density_profile}
    \end{subfigure}%
    ~
    \begin{subfigure}[b]{0.48\textwidth}
        \centering
        \includegraphics[width=\textwidth]{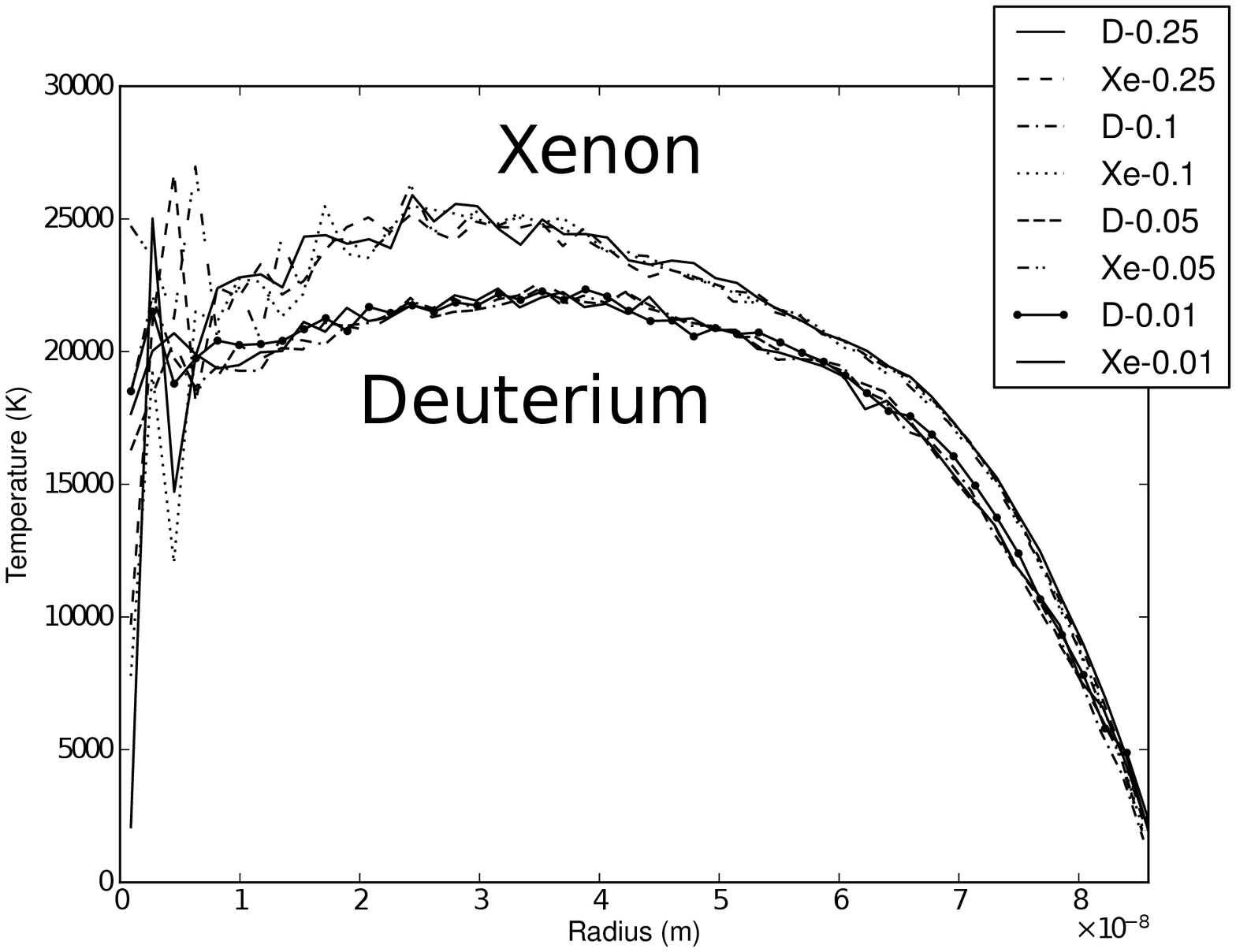}
        \caption{Temperature vs Radius}
        \label{fig:error_temperature_profile}
    \end{subfigure}
    \caption{Density and temperature profiles for multiple error rates at the
        point of shock wave convergence.
        Error rates of 1\%, 5\%, 10\%, and 25\% are shown for each gas present
        in the simulation (deuterium, and xenon).
        The small traces of D$_2$ remaining have been omitted as the low
        concentration results is large, rapid fluctuations its computed
        temperature.
        These examples were generated from a 0.5$\mu$m ambient bubble expanded
        to 20 times its ambient radius and then collapsed with 5 atmospheres of
        constant pressure, in water.
        The gas was originally a 90/10 mixture of xenon and D$_2$, respectively.
        The densities in \ref{fig:error_density_profile} have been normalized
        by the average density of the bubble.}
    \label{fig:error_profiles}
\end{figure*}

Determining the next collision partner for a given particle is necessary for
the event driven algorithm employed here.
In the na\"ive case, determining the next collision events for a given
particle involves an $O(n)$ search by comparing a particle to all potential
collision partners, which gives quadratic time algorithm in the number of
particles in the system.
Fortunately, this can be made more efficient by partitioning the simulation
space, reducing the number of intersection tests to a small constant number by
only considering intersections between particles in the same or adjacent cell.

When the cells are sufficiently small, the cell occupancy will be low and
there will be relatively few intersection tests that need to be performed.
For this optimization to be accurate, however, the edge length of a cell must
not shrink below one atomic diameter (of the largest particles in the
simulation).
The cell partitions are shrunk so as to tightly bound the bubble each time the
volume of the bubble decreases by half or at every snapshot if the cell length
has become less than the expected atomic diameter of the largest gas at room
temperature.
Conversely, if the bubble should expand outside of the partitioned volume, or
if the expected particle diameter becomes too large, then the partitions are
expanded such that either the volume is once again enclosed for the duration
of the snapshot.

Unfortunately, the cell partitions can only be made so small, which greatly
impedes performance when used in the VSS model, where the atomic diameter is
energy dependent -- higher energies results in a lower effective cross
section.
To ameliorate such effects, we introduce an adaptive minimum cell size where
the smallest edge length is determined from the gas in the simulation.
This is particularly effective, as the increasing energy of the system results
in a significantly reduced expected collision cross section.

The collision distance of every particle-particle collision is recorded during
a snapshot.
When it is determined that the cells need resizing, the new minimum edge
length is set to be the $k$-th percentile of the collision cross sections for
all collisions that have occurred during the current snapshot.
The percentile $k$ is variable, and acts as a trade off between performance
and precision.
The error rate for this method is then defined as the complementary
probability $(1 - k)$ and represents the probability that an error could
\emph{potentially} occur.
In general, a collision cross section for a pair of particles larger than the
cell length does not mean that an error will result, as it depends other
possible collisions in the system.
In this case, an error entails a particle colliding with a different particle
than it would if there was no cell partitioning.

In general, this modification only affects the heavier gases in the
simulations, as they tend to have a much larger collision cross section.
It also adversely affects the low energy collisions, as a lower energy results
in a higher collision cross section.
Technically, this potential for error exists regardless of the choice of cell
length, as there will always be a small enough relative velocity that gives
a sufficiently large collision cross section for the VSS model.

In general, the choice of error rate has little effect on the macroscopic
properties of the gas, even during peak temperatures and pressures.
Figure \ref{fig:error_profiles} shows how similar the temperature and density
profiles are for various error rates.
The plots can be unclear due to the high density of the lines in both figures.
The density plots show that the densities for each gas is nearly identical
between the different error rates, and the temperature plots show the same for
temperature.
The two groupings in each plot correspond to deuterium and xenon.

The temperature and densities vary little for most of the bubble's volume.
The only noticeable discrepancies are due to the relatively low volume at the
apex of the cone, where fewer particles give rise to statistical fluctuations
in the recorded temperatures and densities.
This is an artifact of simulating a small portion of the total bubble, not of
the modified partitioning method.

The performance for the error rates presented above range between two and
three times that of the `error free' simulation.
The general procedure is to use a high error rate to search the parameter
space of the simulation in an efficient manner and then rerun potentially
interesting simulations at a lower error rate to validate the results.

As a final consideration, using an adaptive minimum bound in such a way is
potentially more accurate when the gas is cool, as it can increase the minimum
cell size to conform to the conditions of the gas.
Due to thermal wall conditions, the successive rebounds of the KM equation
actually occur with a relatively low gas temperature, which results in the
simulation increasing the minimum cell size; therefore, this modification goes
beyond improving the performance of the simulation when using VSS parameters,
but provides a way to effectively perform spatial partitioning when using an
energy dependent collision cross section model with controllable error bounds.
This is particularly necessary when the volume of the system changes by
several orders of magnitude throughout the duration of the simulation.

\subsection{Event Calendar}
\label{subsec:event_calendar}

Due to the lack of long range forces, there is no need to progressively
integrate the system using a small time step.
Instead, the system can be advanced by the important events in the simulation,
whose occurrences can be determined efficiently and exactly.
The algorithm for doing so is described by \cite{rapaport_event_1980,
rapaport_art_2004}.
Here, the major events in the simulation are particle-bubble collisions,
particle-particle collisions, cell partition crossings, cone crossings, and
system updates (an implementation detail used primarily for book keeping and
managing the solution to the KM equation).

At the start of the simulation, all the events in the simulation are found and
recorded in the event calendar -- a priority queue which also associates
events that involve the same particles.
To advance the simulation, the next event in the calendar (i.e.\ the event
that will occur the soonest) is extracted; all events associated with the
event being extracted are removed from the queue as they are no longer valid.
The simulation then performs the appropriate actions to execute that event,
and all events for the particles involved in the event (up to two due to
particle-particle collisions) are recognized and recorded in the event
calendar.

The event calendar is implemented as a binary tree structure whose ordering is
based on the time of the events in the calendar.
Each tree node also has the pointers required to be present in up to two
circular, doubly-linked lists, which associate all events of the same
particle.
Under the assumption that the event times are randomly distributed, the
expected depth of the tree is $2 \,\textrm{ln}\, N$ based on the analysis by
\cite{knuth_art_1998}.
This may be a spurious assumption during the final moments of collapse due to
the presence of density and temperature heterogeneities, but a detailed study
has not been performed.

It is worth noting that the event calendar is where the majority of processing
time is spent in any simulation consisting of more than a few thousand
particles.
Thanks to the spatial partitioning refinements described in Section
\ref{subsec:spatial_partitioning}, the number of events in the simulation can
be kept to within a small constant factor of the number of particles.
However, in the presence of strong density variations, cell occupancy for some
cells can increase significantly.
This results in not only an increased number of intersection calculations but
also of the number of events in the calendar, as the number of collision
events per particle increases in the high density region.
More events increases the depth of the tree and the number of insertions, but
the large number of associated events also results in a large number of
deletions.
It is often worth spending a non-trivial amount of computational effort to
reduce the number of events in the calendar, even at the cost of more
frequently adjusting the spatial partitions and rebuilding the event calendar.

\subsection{Gas Statistics}
\label{subsec:gas_statistics}

From the available, low level description of the bubble's interior, high level
properties of the gas must be derived.
Of particular interest are the temperature and density of the gas and how they
vary throughout the bubble.
Under the assumption of radial symmetry, we divide the bubble into $N$
non-overlapping shells.
For each shell, we compute the ionization, temperature, and density each each
gas species contained therein as described in \citet{ruuth_molecular_2002}.
The difference here is that we are now considering the possibility of many gas
species inhabiting a given shell.

The temperature of gas $g$ is computed based on the kinetic energy of the gas
in the shell.
Because there is strong radial motion in the gas, the energy due to the radial
motion of the shell is subtracted as follows
\[
    T_g = \frac{m_g}{3 N_g k} \sum_i^{N_g} (v_{g,i}^2 - \bar{v}_g^2),
\]
where $\bar{v}_g$ is the average radial velocity of gas $g$.
The average radial velocity for a collection of particles can be computed as
\[
    \bar{v} = \frac{1}{N} \sum_i^{N} \textrm{proj}_{r_i} v_i,
\]
with $r_i$ being the position of particle $i$.

The maximum collision energy for each gas species pair is also recorded for
the duration of the simulation.
Collision energy is used as another possible metric for how energetic the
system is, especially since collision energy determines whether ionization and
dissociation will occur.
Temperature becomes less reliable when there is strong motion in the bubble
due to shock waves, especially during the final focusing of the shock wave
where it converges at the center of the bubble.
This can result in the averages collision energy exceeding what would be
expected for a given gas temperature, as the radial motion at each shell
partition is subtracted from the total kinetic energy to estimate the
temperature at each shell.

\section{Results}
\label{sec:results}

This section will outline the results of various simulations in an attempt to
give an understanding of how various features affect the evolution of the MD
system.
Previous works have considered the effects of various modeling choices affect
the gas inside of the bubble \citep{BassPhd, bass_molecular_2008,
ruuth_molecular_2002, bass_symmetry_2008}.
These studies are necessarily incomplete without considering how these choices
also affect the evolution of the bubble wall, whose motion has a strong,
quantifiable effect on the gas.
In particular, we will consider how modeling choices and subsequent
modifications made to the simulation affect the results of previous work and
the validity of assumptions made in previous modeling efforts.

In the simulations presented, the bubbles are specified by their composition
and ambient bubble radius, factoring in the surface tension of the liquid and
ambient pressure.
Rather than expand the bubbles using the rarefaction phase of a sinusoidal
pressure function, we simply expand the bubbles by a set factor; the expansion
factors are chosen so as to be physically plausible and are typically less
than what would be experienced using a sinusoid to expand the bubbles.
This is done to save time and minimize potential sources of discrepancies in
timing when comparing simulations.
Simulating the full expansion and collapse of a bubble is time consuming
which, due to the coupling of the bubble wall with the internal molecular
dynamics, is the only way to determine the true maximum radius of the bubble
when using the pressure gauge.
This is due to how the pressure gauge coupled with the various wall methods
results in different temperature gases at the point of maximum expansion.
Therefore, the bubbles would not all have the same starting point (initial
radius), which makes comparison more difficult.

The simulations presented here are of bubble with an ambient radius of
0.5$\mu$m, expanded to 20 times its ambient radius, and then collapsed with
5 atmospheres of constant pressure.
This process occurs in water with a temperature of 300K.
The specifics will vary by section as various features are turned on and off
to facilitate analysis of their effects.
Unless otherwise stated, the simulations use the specular wall conditions and
the VSS molecular model with ionization and dissociation enabled.

\subsection{Pressure Gauge and Gas Composition}
\label{subsec:pressure_gauge}

\begin{figure*}
    \centering
    \begin{subfigure}[b]{0.48\linewidth}
        \includegraphics[width=\linewidth]{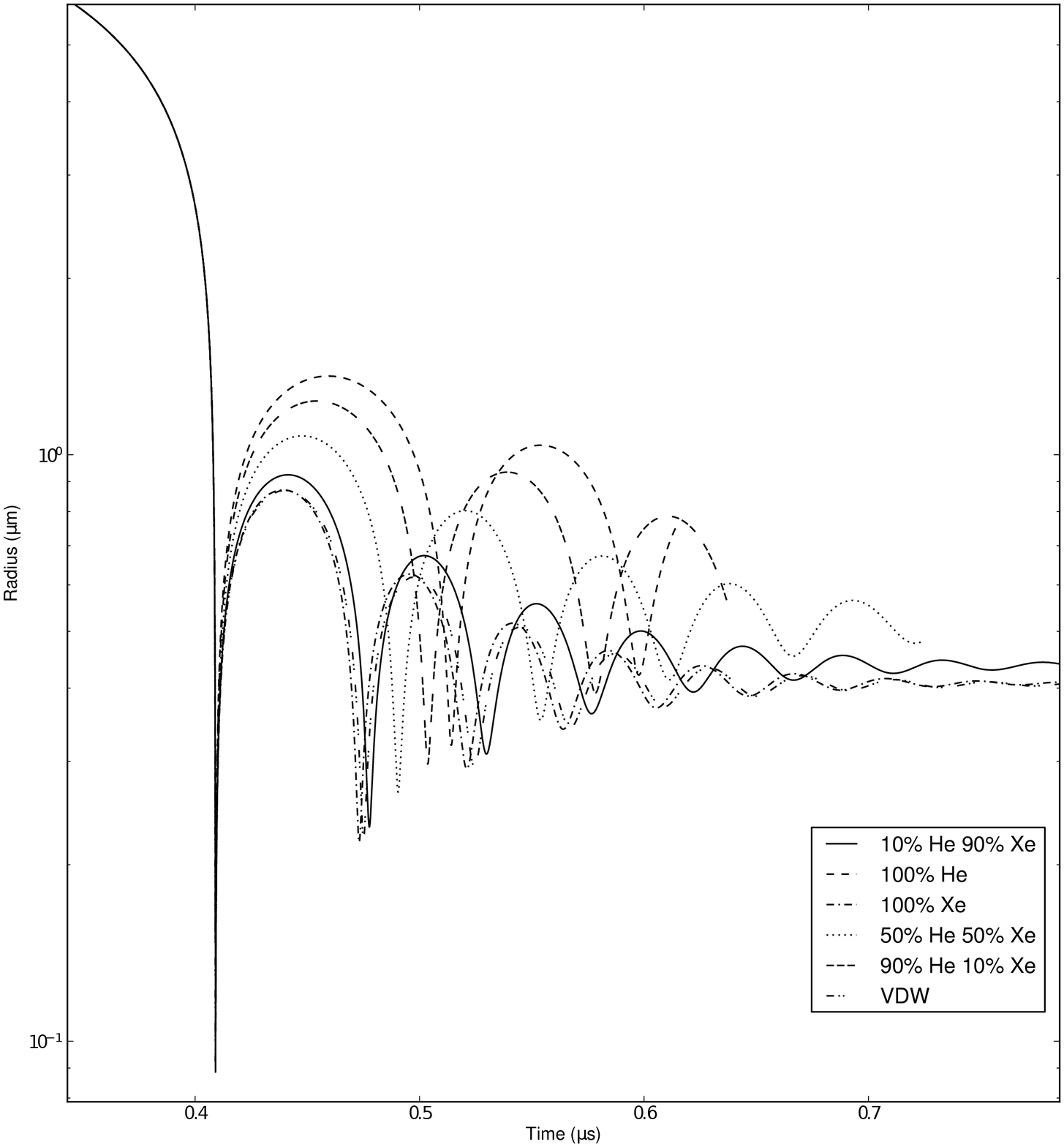}
        \caption{Radius vs time}
        \label{fig:gas_composition_radius}
    \end{subfigure}%
    ~
    \begin{subfigure}[b]{0.48\linewidth}
        \centering
        \includegraphics[width=\linewidth]{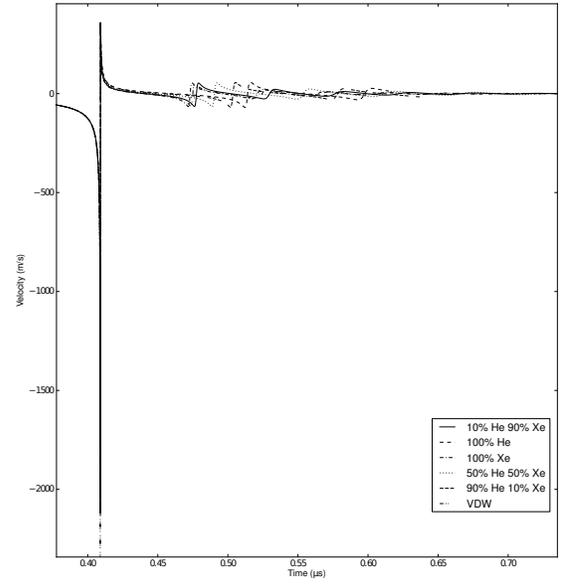}
        \caption{Velocity vs time}
        \label{fig:gas_composition_velocity}
    \end{subfigure}%
    \\
    \begin{subfigure}[b]{0.48\textwidth}
        \centering
        \includegraphics[width=\linewidth]{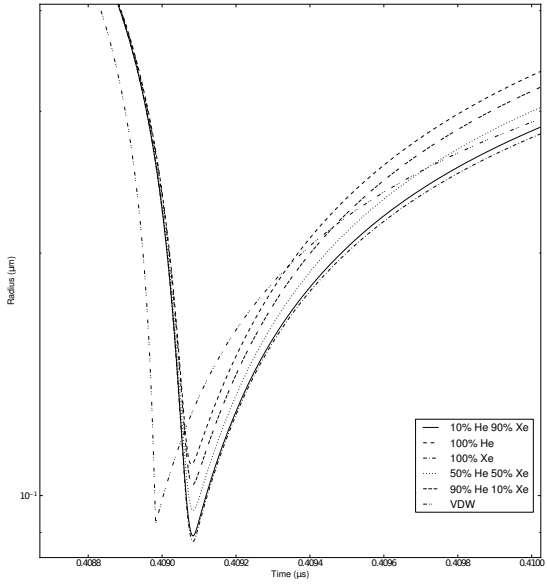}
        \caption{Expansion of radius vs time}
        \label{fig:gas_composition_radius_zoom}
    \end{subfigure}%
    ~
    \begin{subfigure}[b]{0.48\textwidth}
        \centering
        \includegraphics[width=\linewidth]{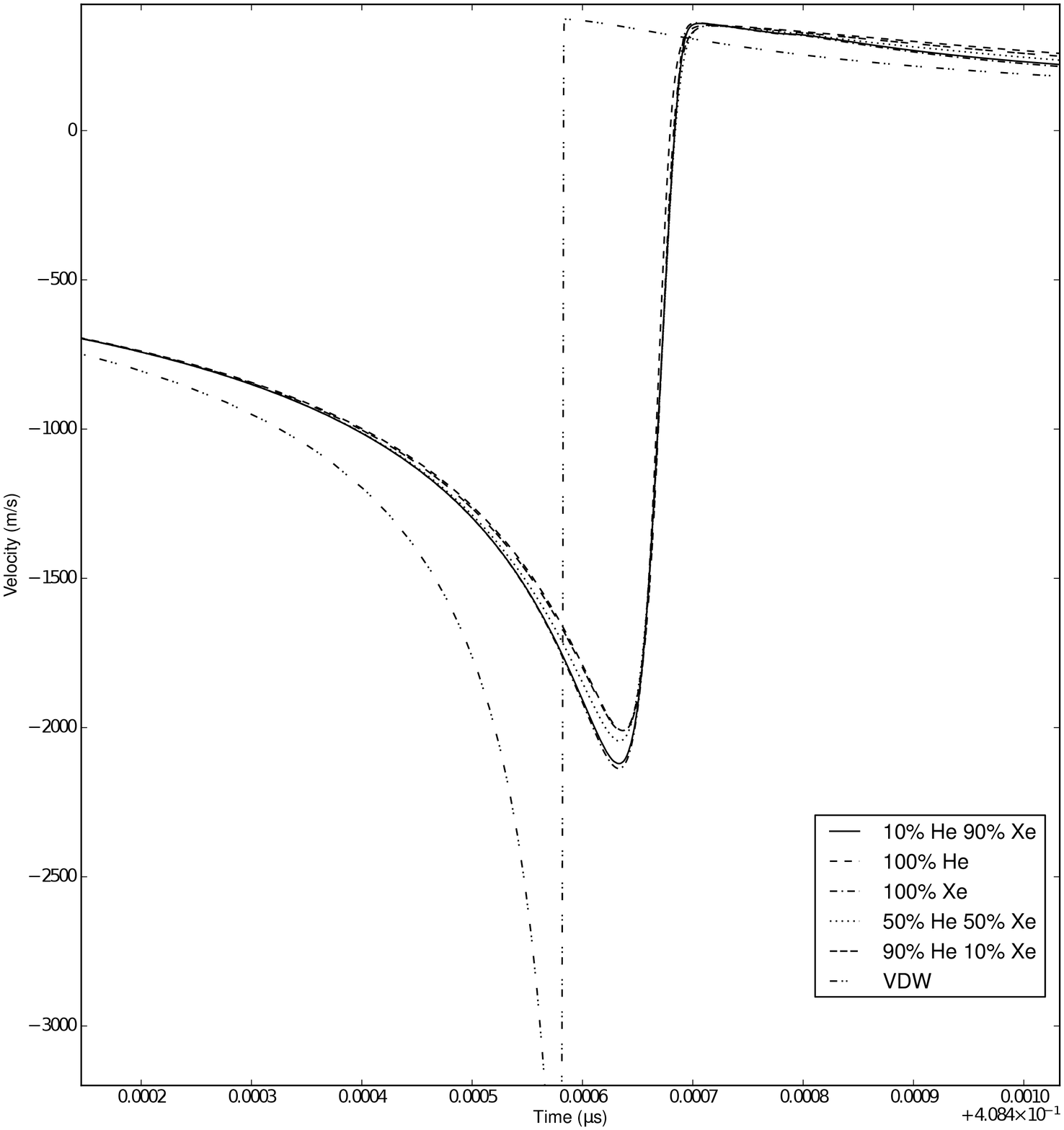}
        \caption{Expansion of velocity vs time}
        \label{fig:gas_composition_velocity_zoom}
    \end{subfigure}
    \caption{Radius versus time curve during the collapse and damped
        oscillations of a bubble with differing gas compositions
        (but are otherwise the same).
        The analytical solution to the KM equation for the same parameters are
        included as well, for comparison. The expansions show the time
        surrounding the minimum radius of the bubble, when densities and
        temperatures are greatest.}
    \label{fig:pressure_gauge_radius_and_velocity}
\end{figure*}

\begin{figure*}[t]
    \begin{subfigure}[b]{0.48\textwidth}
        \centering
        \includegraphics[width=\linewidth]{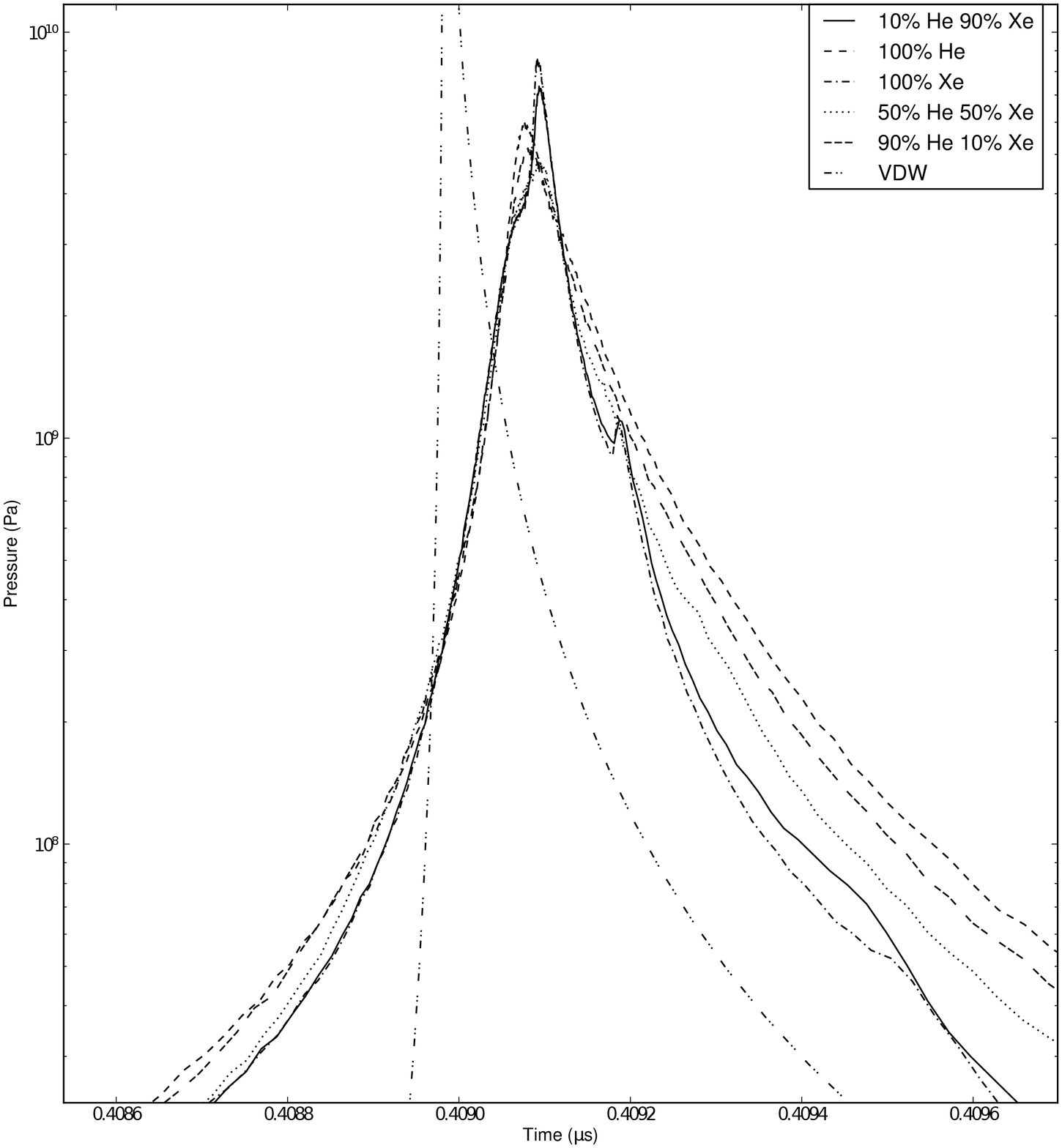}
        \caption{Pressure vs time for specular wall conditions}
        \label{fig:gas_composition_pressure_specular}
    \end{subfigure}%
    ~
    \begin{subfigure}[b]{0.48\textwidth}
        \centering
        \includegraphics[width=\linewidth]{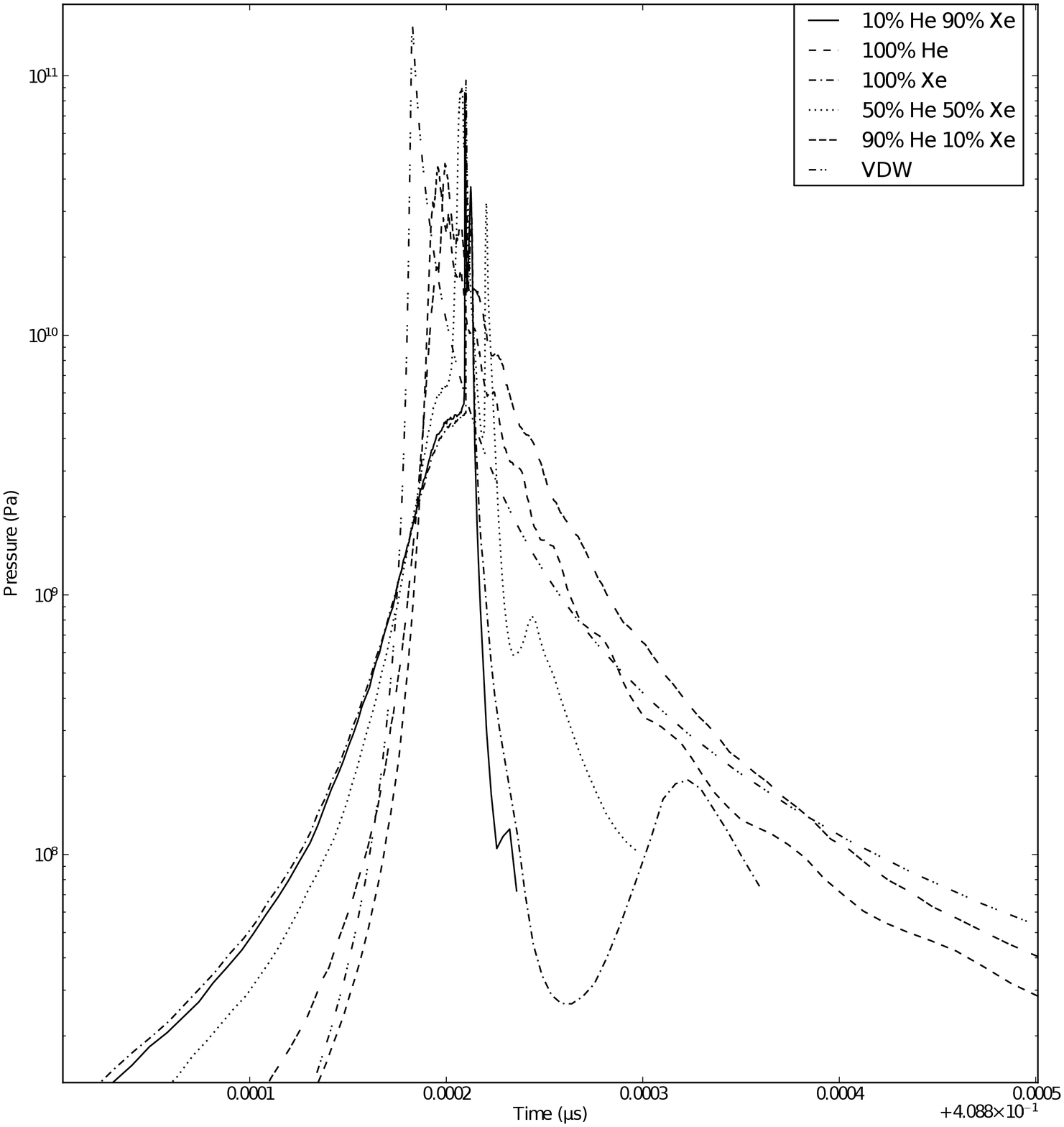}
        \caption{Pressure vs time for thermal wall conditions}
        \label{fig:gas_composition_pressure_thermal}
    \end{subfigure}
    \caption{Pressure versus time plots for specular and thermal wall
        conditions.
        The simulation parameters are the same as in Figure
        \ref{fig:pressure_gauge_radius_and_velocity}, and show the pressure as
        measured around the initial rebound of the bubble for the two wall
        conditions.}
    \label{fig:pressure_gauge_pressure}
\end{figure*}

The inclusion of gas pressure has a marked impact on the evolution of the
bubble wall through time.
The strong density variations during the collapse of the bubble result in
a deviation from both adiabatic and VDW equations of state, some of
which is due to modeling assumptions and the properties of the gas in the
system.
The pure xenon bubble deviates least from the VDW based solution, which does
not take into account properties specific to each gas species.

Figure \ref{fig:gas_composition_radius} illustrates the impact that the
inclusion of the pressure gauge has on the evolution of the bubble.
The KM equation coupled with the empirical pressure experiences less
acceleration as its velocity increases, due to the non-negligible motion of
the bubble wall relative to the gas and, in some of the gas compositions,
formation of a shock wave, which results in a region of high density gas
exerting additional pressure on the bubble wall.
Despite the initial slowing, the minimum radius for the empirical pressure is
less than that of the VDW pressure in some cases.
This is heavily dependent upon the simulation parameters (e.g.\ compare the
values in Tables \ref{tab:rpk_gas_composition} and
\ref{tab:rpk_gas_composition_thermal_wall}).

Figure \ref{fig:pressure_gauge_pressure} shows just how the empirically
determined pressure deviates from the VDW pressure for the two wall
conditions.
While there is some difference before the rebound of the bubble, the major
discrepancies occur at the rebound and right after.
For the empirically determined pressures, there is a consistent rise in
pressure as the bubble collapses, which is higher than the adiabatic pressure,
due to the shock wave forming at the bubble wall.
In heavier gases, prior to the rebound, the rate of pressure increase will
drop for a short period of time, when the shock wave detaches itself from the
bubble wall; the duration of this pause is a function of the speed of sound in
the different gas mixtures.
These shock waves are more prominent in the thermalizing wall and can be seen
in their influence upon the pressure recorded at the bubble wall, an effect
that can be seen by comparing Figures
\ref{fig:gas_composition_pressure_specular} and
\ref{fig:gas_composition_pressure_thermal} Pressure suddenly spikes after the
pause, as the shock wave rebounds, which is what causes the bubble wall to
finally rebound.

The effects of gas composition and wall conditions are difficult to completely
separate.
The thermalizing wall strengthens the shock profiles in the gases, which would
only be present in the specular wall for the heaviest gases.
The effects of which can be seen in the pressure gauge readings for the two
wall methods.
The heaviest gases display strong shock profiles which traverse the bubble many
times after the initial collapse in the thermal wall.
Even light gases display small deviations due to shock formation under the
thermal wall, even though no shocks are present in the specular wall for the
light gases.
Further comparison of the wall methods is carried out in Section
\ref{subsec:wall_condition_effects}.

After the rebound, the pressure gradually decreases, but with a sinusoidal
component overlaid when strong shocks are present.
The sinusoidal component is due to the continued propagation of the shock wave
through the gas, which gradually decreases in strength and frequency as the
bubble expands.
This effect is most prominent in the heavier gases present in Figure
\ref{fig:gas_composition_pressure_thermal}.
In all cases, heavier gases have a longer period and higher amplitude.

\begin{table}
    \centering
    \scriptsize
    \begin{tabular}{cc|cccc}
        \hline \hline
        \multicolumn{2}{c|}{Composition} & $R_{min}$ & $V_{\textrm{max}}$
            & $T_\textrm{max}$ & $E_\textrm{max}$ \\
        \% He & \% Xe                        & ($\times 10^{-8}$m) & (km/s)
            & (K)      & (eV) \\ \hline

        100 & 0   & 10.93  & 2.010 & 135052 & 107  \\
        90  & 10  & 10.32  & 2.014 & 99826  & 82.8 \\
        50  & 50  & 9.585  & 2.045 & 128443 & 66.1 \\
        10  & 90  & 8.914  & 2.121 & 90200  & 41.9 \\
        0   & 100 & 8.774  & 2.138 & 95607  & 57.7 \\ \hline

        \multicolumn{2}{c|}{VDW}   & 9.2455 & 3.918   & n/a & n/a \\
        \hline \hline
    \end{tabular}
    \caption{Bubble and gas statistics for different internal gas compositions.
        The maximum temperature and collision energies are for the lightest gas
        in the simulation.
        For reference, the analytical solution for the VDW solution to the KM
        equations are also included.
        These values are taken from the simulations presented in
        \ref{fig:pressure_gauge_radius_and_velocity}.}
    \label{tab:rpk_gas_composition}
\end{table}

These results illustrate not only the importance of gas composition with
respect to the bubble wall dynamics, but also illustrate effects that shock
wave formation has on the motion of the bubble wall.
The shock waves in the results presented in this section are relatively weak.
The same bubbles using the thermalizing wall results in stronger shock profiles.
As pointed out by \citet{prosperetti_modelling_1999}, shocks in the gas are
frequently ignored as they are not necessary to achieve the temperatures
sufficient to produce light emission.
However, we find that shock formation has a strong influence on the motion of
the bubble wall and upon the internal temperatures of the bubble, even if
shock formation is not necessary to explain sonoluminescence.
Furthermore, this shock wave has a strong influence on the collision energies
and temperatures at the center of the bubble, which may have ramifications for
fusion.

\subsection{Pressure Gauge and Sphere Dynamics}
\label{subsec:pressure_gauge_sphere_dynamics}

\begin{figure}[t]
    \centering
    \includegraphics[width=\linewidth]{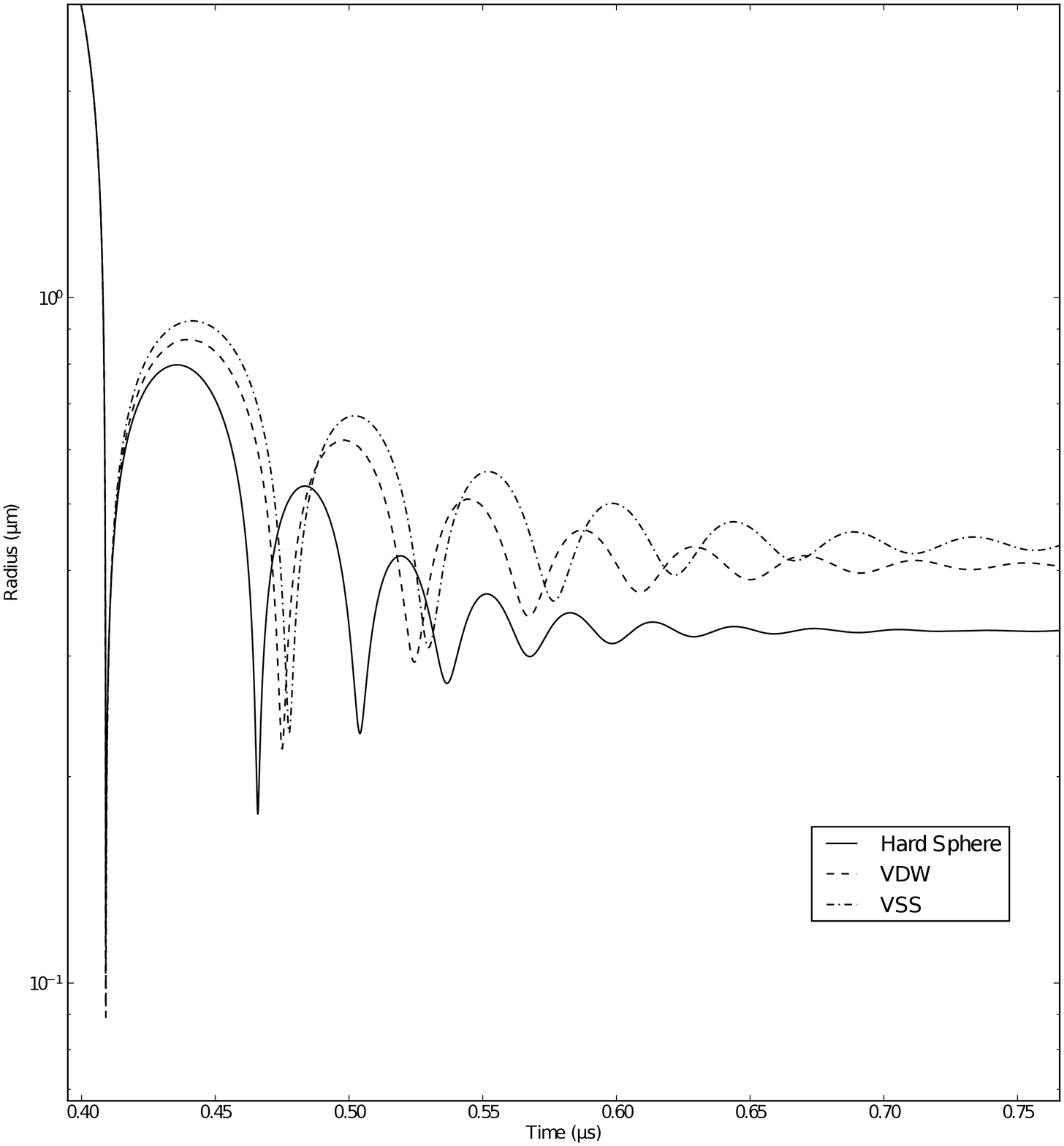}
    \caption{Radius vs time curve for different collision cross section modes.
        Included are with both a constant diameter and the variable soft
        sphere models.
        For reference, the KM equations using adiabatic pressure and VDW hard
        core pressure are included.
        The only difference between simulations was the choice of diameter model
        within the simulation.
        The KM curve with van der Waal's hard core pressure is
        included for reference.
        The gas composition of the bubble was 90\% xenon and 10\% Helium.}
    \label{fig:sphere_type_rpk}
\end{figure}

\begin{table}[t]
    \centering
    \scriptsize
    \begin{tabular}{c|cccc}
        \hline \hline
        Molecular   & $R_\textrm{min}$    & $V_\textrm{max}$ & $T_\textrm{max}$
            & $E_\textrm{max}$ \\
        Model       & ($\times 10^{-8}$m) & (km/s)           & (K)
            & (eV)          \\ \hline
        Hard Sphere & 11.795              & 2.243            & 69474
            & 38.1          \\
        VSS         & 8.908               & 2.121            & 86666
            & 39.8          \\
        \hline \hline
    \end{tabular}
    \caption{Comparison of bubble statistics for hard sphere and VSS molecular
        model simulations.
        The data is taken from the simulations used to generate Figure
        \ref{fig:sphere_type_rpk}.}
    \label{tab:sphere_type}
\end{table}

Previous work \citep{ruuth_molecular_2002} found that molecular models will
frequently produce similar results.
The major difference between the constant diameter and VSS model was found to
be its effect on ionization due to different atomic diameters affecting the
collision probability.
This, in turn, affected the overall temperature of the gas, as less energy was
lost ionizing the constituent particles.
These results ignore one of the most important effects of changing the
diameter of the particles in the simulation; the molecular model most affects
the volume of space occupied by the gas particles within the simulation.

With the inclusion of the pressure gauge, we find that the average level of
ionization within the bubble is not in keeping with the results of
\citet{ruuth_molecular_2002} for the simulations presented here.
Immediately after the point of minimum radius, the average ionization level
for the hard sphere simulation is 1.69 ionizations for xenon and 0.065 for
helium while the VSS simulation has an average ionization of 1.78 for
xenon and 0.16 for helium.
This is due to the higher compression in the VSS simulation, which resulted in
a longer, higher density period where the gas particles could ionize.
Such an effect would not be present using the VDW estimate of pressure, as
the wall motion is then independent of the molecular model.
This differences in relevant bubble conditions can be seen in Table
\ref{tab:sphere_type}.

Even though these results do not agree on how the molecular model affect the
level of ionization within the bubble, they do show the same trend in
temperature.
The VSS model still has a higher peak temperature, but part of that is due to
the greater compression of the bubble contents, rather than less energy being
lost to ionization.

The VSS model gives a decreasing atomic diameter with respect to the energy in
the system.
This decreases the theoretical minimum bubble volume as a result, which
becomes a major factor during the high density and temperature conditions
experienced within the bubble and contributes to the VSS bubble being able to
compress lower minimum radius than the hard sphere bubble.
Figure \ref{fig:sphere_type_rpk} illustrates the effect that the choice of
molecular model has on the evolution of the bubble.
The constant diameter model most closely approximates the VDW pressure curve,
as the choice of hard core radius is chosen to be close to the packing density
of hard spheres for effective atomic diameters at room temperature.

The choice of gas properties, therefore, has a strong influence on the
evolution of the bubble, which can alter temperatures and densities within the
bubble in a manner that is not readily detectable without an empirical
estimation of pressure.
This also illustrates the importance of using an accurate model of atomic
diameter in the simulation.
It has been noted by \citet{bass_molecular_2008} that the VSS energy-diameter
dependence is stronger than is expected for the Lennard-Jones 6-12 potential,
and that it diverges at high temperatures \citep{fan_generalized_2002}.
The constant diameter model and VSS model, therefore, provide limiting cases
for what is physically correct.

\subsection{Wall Condition Effects}
\label{subsec:wall_condition_effects}

\begin{figure*}[t]
    \begin{subfigure}[b]{0.48\textwidth}
        \centering
        \includegraphics[width=\linewidth]{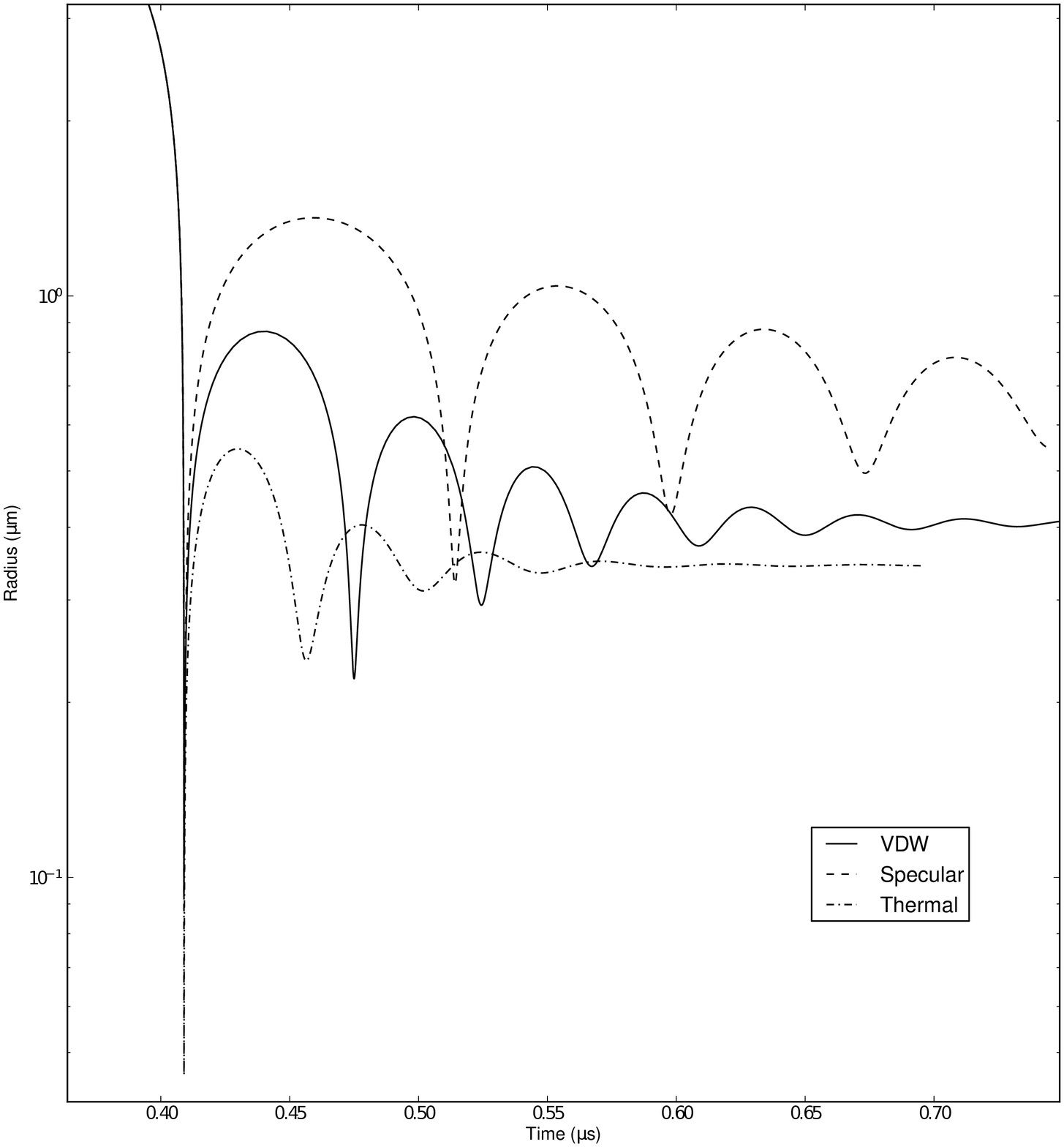}
        \caption{Radius vs time}
        \label{fig:wall_comparison_radius}
    \end{subfigure}%
    ~
    \begin{subfigure}[b]{0.48\textwidth}
        \centering
        \includegraphics[width=\linewidth]{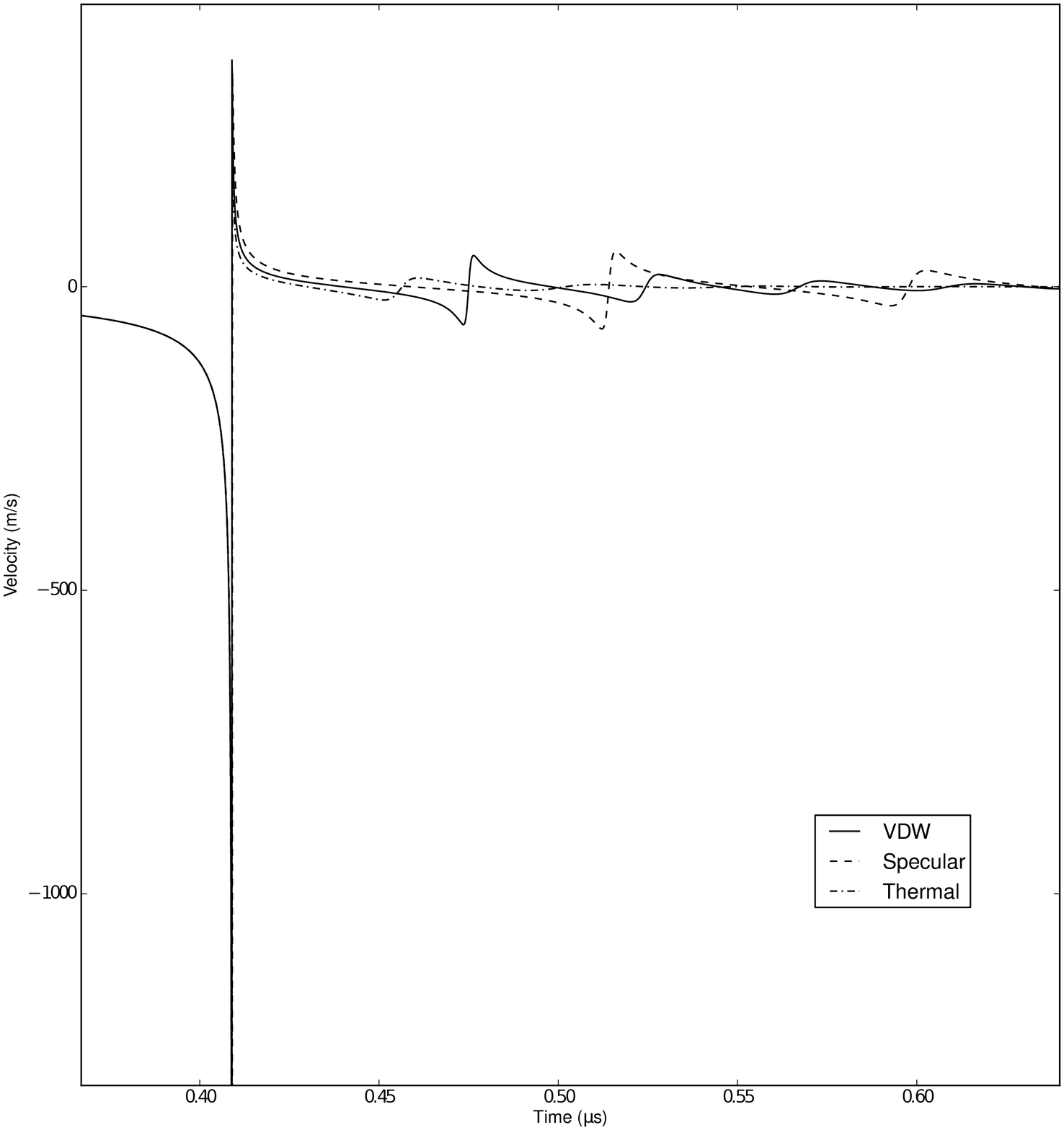}
        \caption{Velocity vs time}
        \label{fig:wall_comparison_velocity}
    \end{subfigure}
    \caption{Radius and velocity versus time plot of the bubble under the two
        different wall conditions.
        These simulations were run with pure helium bubbles.
        The only between the two simulations is the boundary conditions employed
        at the bubble wall.
        For comparison, the numerical solution to the KM equation using VDW hard
        core pressure is also included.}
    \label{fig:wall_comparison_radius_and_velocity}
\end{figure*}

\begin{figure}[t]
    \centering
    \includegraphics[width=\linewidth]{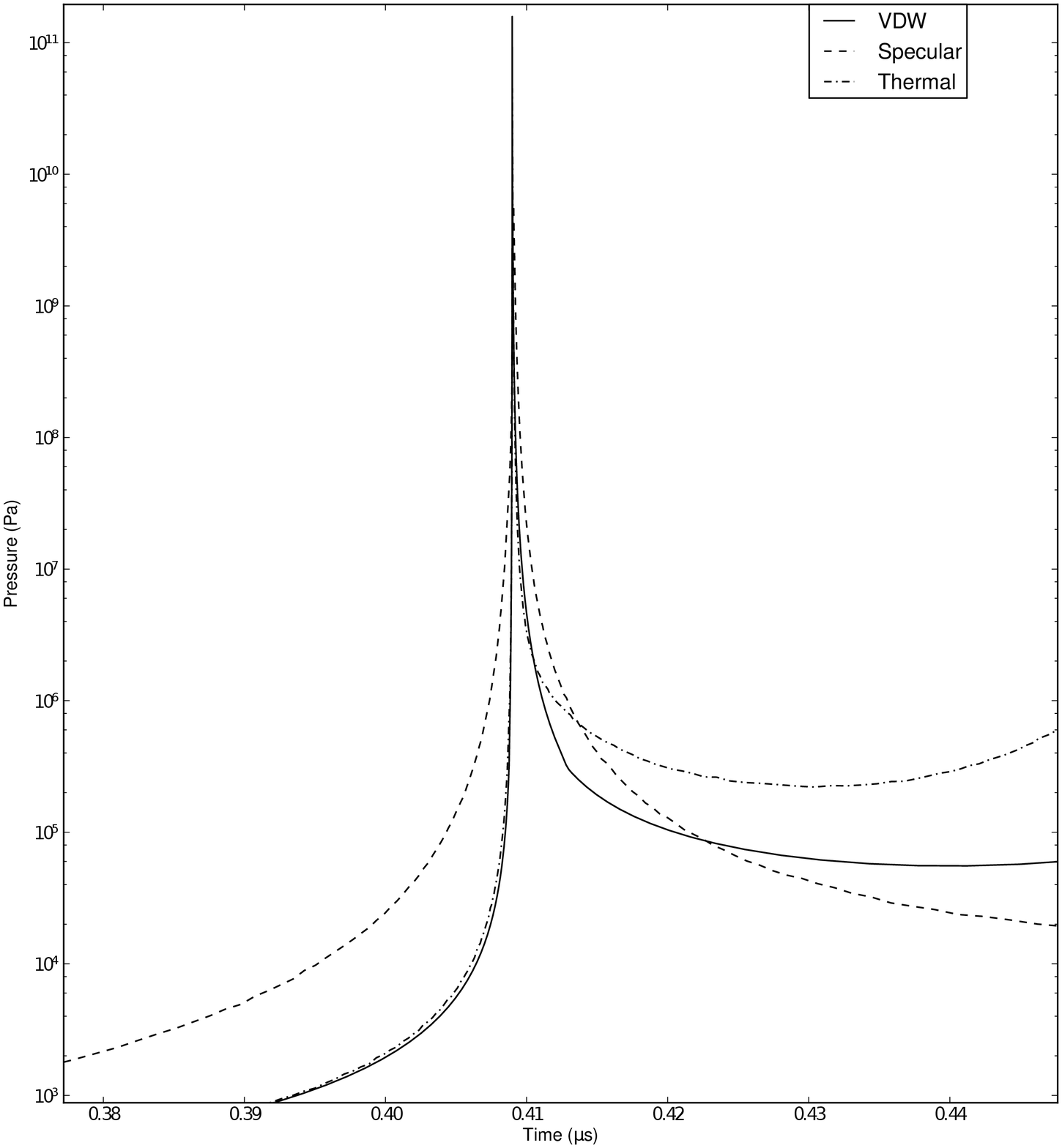}
    \caption{Pressure vs time plot for the same bubble as Figure
        \ref{fig:wall_comparison_radius_and_velocity} but restricted to the
        time around the rebound of the bubble.}
    \label{fig:wall_comparison_pressure}
\end{figure}

Previous work has shown that heat exchange between the bubble wall results in
a cooling of the bubble for most of its compression cycle, but can lead to
stronger shock profiles and higher peak temperatures, due to increased energy
focusing \citep{ruuth_molecular_2002}.
While the results here find that the thermalizing boundary does strengthen the
shock profile, the additional energy focusing characteristics of the shock wave
are not always sufficient to make up for the energy lost to the thermalizing
wall.
This is likely due to various changes in the simulation parameters used here,
so these results may differ.
Though a detailed study of the divergence has not been conducted, we suspect
this is due to their use of an undamped Rayleigh-Plessest style equation
results in significantly higher wall velocities, yielding much strong shock
waves.
In addition, we consider a pre-expanded bubble subjected to a constant driving
pressure, rather than a bubble expanded by a sinusoidal pressure function.
Our expansion factor of twenty times is less than what would be expected from
even a 1.5 bar sinusoidal driver, which will result in a lower collapse
velocity as well.

An important aspect to consider is how these boundary conditions affect the
motion of the bubble wall.
We have demonstrated that the gas dynamics can have a large impact on the
evolution of the bubble wall, but done using the same wall conditions for all
gases.
Because the pressure is computed from the change in momentum of particles
striking the bubble wall, the method used to compute the outgoing vector for
particles impacting the wall has a strong influence on the evolution of the
bubble wall.
The wall conditions determine how work done by the driving force is distributed
between the kinetic energy of the bubble wall and the gas particles.
The boundary conditions also have a clear effect on the gas dynamics, so it is
important to consider how this may interplay with the motion of the bubble
wall.
The gas will lose a non-negligible amount of energy to the wall during its
collapse using the thermalizing wall, influencing temperatures within the
bubble, altering ionization levels, and promoting shock formation.

\begin{table}[t]
    \centering
    \scriptsize
    \begin{tabular}{c|cccc}
        \hline \hline
        Wall       & $R_\textrm{min}$    & $V_\textrm{max}$ & $T_\textrm{max}$
            & $E_\textrm{max}$ \\
        Conditions & ($\times 10^{-8}$m) & (km/s)           & (K)
            & (eV)          \\ \hline
        Specular   & 10.945              & 2.006            & 122487
            & 108.7         \\
        Thermal    & 4.572               & 5.091            & 51005
            & 53.0          \\
        VDW        & 9.246               & 3.918            & n/a
            & n/a           \\
        \hline \hline
    \end{tabular}
    \caption{Comparison of bubble statistics for specular and thermal
        wall conditions. The bubble parameters are the same as those in
        Figure \ref{fig:wall_comparison_radius}.}
    \label{tab:wall_comparison}
\end{table}

\begin{table}[t]
    \centering
    \scriptsize
    \begin{tabular}{cc|cccc}
        \hline \hline
        \multicolumn{2}{c|}{Composition} & $R_{min}$ & $V_{\textrm{max}}$
            & $T_\textrm{max}$ & $E_\textrm{max}$ \\
        \% He & \% Xe                        & ($\times 10^{-8}$m) & (km/s)
            & (K)      & (eV) \\ \hline

        100 & 0   & 4.569 & 5.093 & 61388  & 51.3 \\
        90  & 10  & 4.993 & 4.369 & 91703  & 50.3 \\
        50  & 50  & 5.572 & 3.400 & 56592  & 55.0 \\
        10  & 90  & 6.227 & 2.996 & 76898  & 70.2 \\
        0   & 100 & 6.439 & 2.924 & 111870 & 68.9 \\ \hline

        \multicolumn{2}{c|}{VDW}   & 9.2455 & 3.918   & n/a & n/a \\
        \hline \hline
    \end{tabular}
    \caption{Table of bubble and gas statistics for the same bubble
        configurations as in Table \ref{tab:rpk_gas_composition} but using the
        thermalizing bubble wall rather than specular.
        Note the reversal of trends for the minimum radius and maximum velocity
        when compared to the specular results.}
    \label{tab:rpk_gas_composition_thermal_wall}
\end{table}

Figures \ref{fig:wall_comparison_radius_and_velocity} and
\ref{fig:wall_comparison_pressure} show what a drastic effect the choice
of boundary conditions can have on the evolution of the bubble wall itself.
Because the thermalizing wall siphons energy from the gas, pressure initially
builds at a much slower rate
This causes the bubble wall to collapse more quickly and results in
a significantly smaller minimum radius for the bubble (see Table
\ref{tab:wall_comparison}).
That said, the smaller radius and higher wall velocities are attained only due
to the loss of energy in the system, and that the improved energy focusing and
strong shocks are not sufficient to make up for the total loss in energy of
the system.

The differences presented in Table \ref{tab:wall_comparison} will likely
differ significantly based on the choice of gas and other simulation
parameters.
As shown in Section \ref{subsec:pressure_gauge}, gas composition has a large
effect on the formation of shock waves within the system.
The bubble contents of the simulations presented in this section are pure
helium, which does not experience shocks as strongly as other gases, due to
its high speed of sound.
Strong shocks may also interplay strongly with how the wall conditions
influence energy loss within the system.

The loss of energy to the system is most noticeable during the subsequent
oscillations of the bubble.
Specular reflection at the bubble wall results in a markedly strong rebound
and higher amplitude, lower frequency oscillations of the bubble.
The much stronger rebounds are to be expected of the specular wall, as no
energy is lost to the surrounding fluid during collapse.
It is worth noting that the specular and thermal wall conditions are limiting
cases of the physically realistic wall conditions.
For the conditions simulated here, the specular and thermal wall conditions
bound the analytical solution with VDW hard core pressure.

As shown in \citet{lofstedt_toward_1993}, a first order accurate RP equation
coupled with VDW equations of state yields a radius versus time curve that is
in strong agreement with experimental results.
Their work considers a 10.5$\mu$m ambient bubble subjected to a 1.075
atmosphere acoustic driver at 26.5kHz.
A visual inspection shows the evolution of the bubble to have subsequent
oscillations of higher amplitude and lower frequency, though they do not
differ as strongly as the specular reflection simulations differ from the
idealized VDW adiabatic pressure.
Additionally, \citet{kim_validation_2008} concluded that the collapse of a
sonoluminescing bubble ``is characterized by an almost adiabatic process
even though a large amount of heat transfer through the bubble wall.''
This is a strong indication that the true boundary conditions lie somewhere
between the two methods and merits further study.
Fortunately, the coupling of the bubble wall with the gas dynamics gives
a potential avenue for validation in finding a more suitable model.

Finally, a somewhat distressing interplay exists between the VSS molecular model
and the thermal wall conditions presented here.
Because the thermal wall overestimates the cooling effect at the wall, the
effective molecular diameters remain larger than under specular wall
conditions.
While this influences the evolution of the bubble, it can have a noticeable
effect on the performance of the simulation when modeling large atoms such as
xenon, where the cooling just after rebound results in the cell partitions
expanding.
This, in turn, results in an explosion in the number of collision
intersections that must be examined, due to the high density of the system at
that point.
It is not clear how this effect can be ameliorated without using a more accurate
model of heat transfer between the liquid and the gas.

\subsection{Ionization and Dissociation Effects}
\label{subsec:ionization_effects}

\begin{figure}[t]
    \centering
    \includegraphics[width=\linewidth]{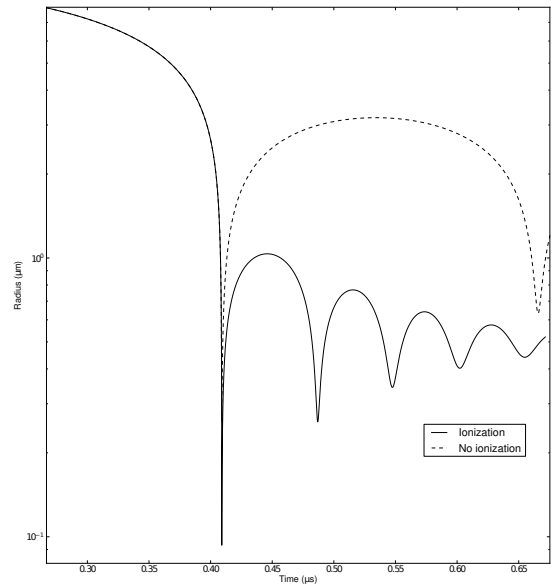}
    \caption{Radius vs time plot comparing the argon simulations with and
        without ionization enabled.}
    \label{fig:ionization_comparison_radius}
\end{figure}

\begin{figure}[t]
    \centering
    \includegraphics[width=\linewidth]{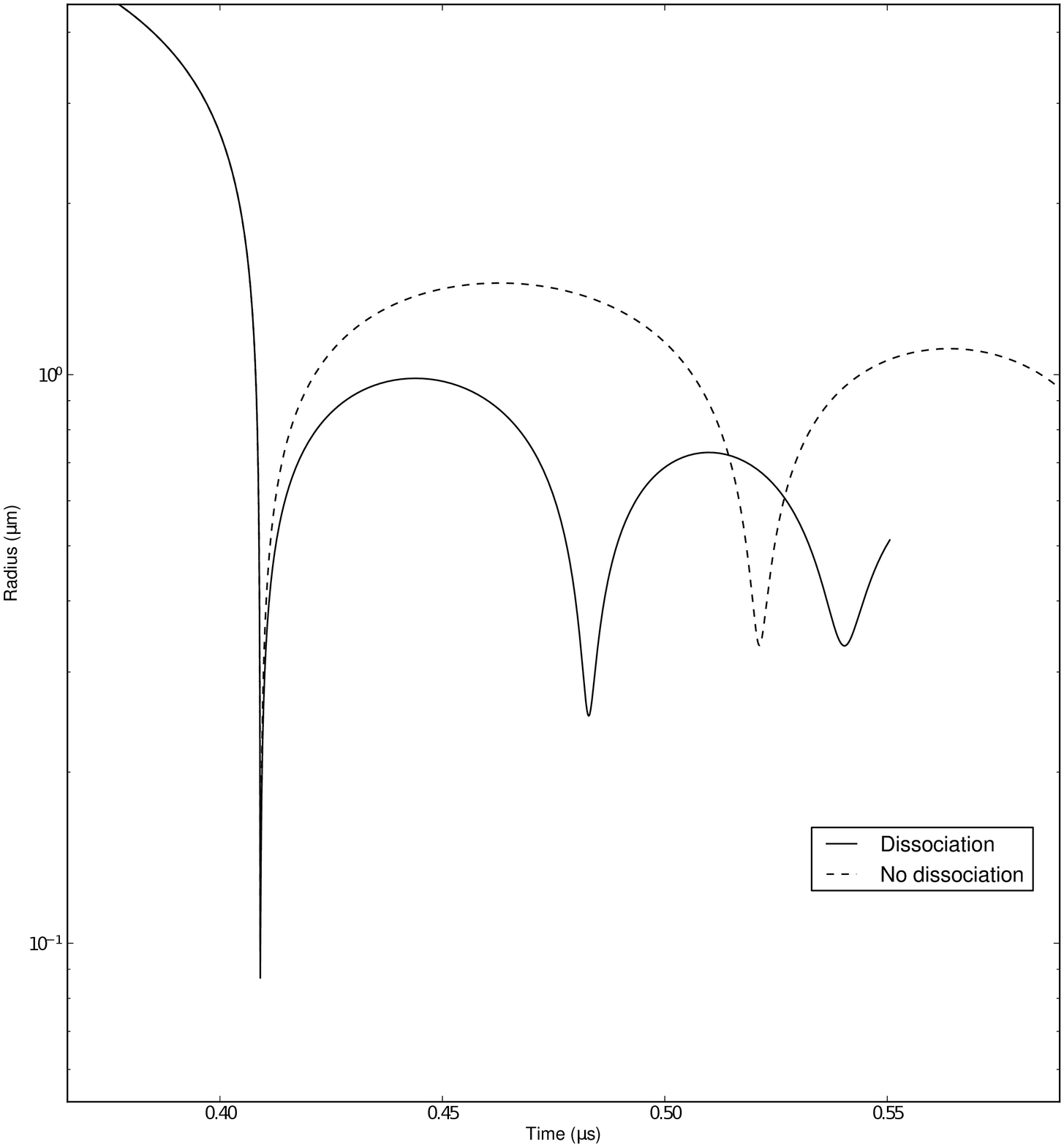}
    \caption{Radius vs time plot comparing deuterium based simulations with and
        without dissociation enabled.}
    \label{fig:dissociation_comparison_radius}
\end{figure}

Ionization has the effect of removing energy from the particle system.
Previous work \citep{ruuth_molecular_2002} mentioned that the only significant
effect of this energy loss was a reduction in temperature.
We find that the inclusion of ionization effects also have a significant
influence on the motion of the bubble wall.
The ionization process removes energy, which results in a drop in the total
pressure of the system.
This is reflected in the motion of the bubble wall in our model.

Due to the different energies required to ionize the different gases, some of
the variation presented in Section \ref{subsec:pressure_gauge} can be
explained through the different amounts of energy lost through ionization to
the different gases.
Here, we present two sets of simulations using the same parameters, but with
different gas compositions.
The simulations were run with helium and argon bubbles, each with ionization
effects turned on and off.
The results for each gas are compared between the two ionization modes to
examine what differences ionization has on the evolution of the bubble wall.

The first property of note is the average ionization level of each bubble.
After the initial collapse, the average ionization level within the ionizing
argon bubble was $\approx 1.32$ ionizations per atom, while the ionizing
helium bubble was $\approx 1.03$ ionizations per atom.
For the argon bubble, this results in a net loss of about 2.53 MJ/mol, while
the helium bubble loses 2.32 MJ/mol (assuming all particles of a given species
do not differ by more than one ionization level).
The bubbles have the same number of gas particles, so these numbers can be
compared directly.

\begin{table}[t]
    \centering
    \scriptsize
    \begin{tabular}{c|cccc}
        \hline \hline
        Simulated       & $R_\textrm{min}$    & $V_\textrm{max}$
            & $T_\textrm{max}$ & $E_\textrm{max}$ \\
        Gas             & ($\times 10^{-8}$m) & (km/s)           & (K)
            & (eV)             \\ \hline
        Helium (No ion) & 25.983              & 1.159            & 660031
            & 830              \\
        Helium (Ion)    & 10.917              & 2.007            & 138430
            & 111              \\
        Argon (No ion)  & 25.988              & 1.158            & 820220
            & 705              \\
        Argon (Ion)     & 9.329               & 2.241            & 86343
            & 69.6             \\
        \hline \hline
    \end{tabular}
    \caption{Comparison of bubble statistics for ionizing and non-ionizing
        simulations.}
    \label{tab:ionization_comparison}
\end{table}

Even though the ionization levels for the two gases require different amounts
of energy, the total amount of energy lost to ionization is very similar.
It is not surprising, then, that the addition of ionization to the two gases
has a very similar effect.
Ionization has a strong cooling effect within these simulations, as the specular
boundary was used.
In specular mode, the energy of the system can only be lost through damping
effects on the bubble wall, ionization, and dissociation.

The loss of energy due to ionization has the additional effect of increasing
the effective collision cross section in the VSS model.
We have already shown (see Section
\ref{subsec:pressure_gauge_sphere_dynamics}) how an altered collision cross
section can have macroscopic effects on the motion of the bubble wall, as the
effective volume occupied by the gas particles is altered.
Based on the results in Table \ref{tab:ionization_comparison}, it is clear that
the VSS model will be strongly influenced by the inclusion of ionization.
The maximum temperature within the non-ionizing argon bubble was over nine times
that of the ionizing argon bubble.
The effects of decreased collision cross section is not enough to overtake the
increased pressure resulting from the high temperatures within the non-ionizing
bubble.

The most noticeable way that the loss of energy influences the system is through
the motion of the bubble wall.
Figure \ref{fig:ionization_comparison_radius} shows the same simulation with
ionization turned on and off.
The ionizing simulation has a less energetic rebound and subsequent oscillations
occur with a lower amplitude and higher frequency.
Indeed, the effects on the oscillations of the bubble due to ionization are
more pronounced than those due to the choice of wall conditions -- though
ionization effectively halts after a certain point is reached, while energy
loss through the wall continues until thermal equilibrium.

Dissociation has a similar effect to that of ionization.
The energy lost to dissociation has the effect of reducing pressure, which
allows the bubble to collapse more rapidly.
Even though the dissociating bubble collapses sooner due to its higher
acceleration when dissociation drops the internal pressure, its maximum wall
velocity is nearly identical to that of the non-dissociating bubble.
The trend for minimum radius between dissociation and non-dissociation is the
same as for ionization and non-ionization, however, the trend for maximum wall
velocity does not correspond between ionization and dissociation.
The exact cause for this is not clear, though it may be caused by the change
in volume occupied by the particles within the bubble.

\begin{table}[t]
    \centering
    \scriptsize
    \begin{tabular}{c|cccc}
        \hline \hline
        Dissociation    & $R_\textrm{min}$    & $V_\textrm{max}$
            & $T_\textrm{max}$ & $E_\textrm{max}$ \\
        Status          & ($\times 10^{-8}$m) & (km/s)           & (K)
            & (eV)             \\ \hline
        No Dissociation & 10.136              & 2.460            & 140590
            & 150              \\
        Dissociation    & 8.681e-08           & 2.438            & 34522
            & 37.3             \\
        \hline \hline
    \end{tabular}
    \caption{Bubble statistics for dissociation effects comparison.
        These simulations were carried out with the same bubble parameters, but
        with dissociation enabled for only one.
        Gas species specific parameters are reported for the smallest gas
        particles in the simulation (D$_2$ for the non-dissociating
        simulations and D for the dissociating simulations).}
    \label{tab:dissociation_comparison}
\end{table}

The deuterium in the simulations for Figure
\ref{fig:dissociation_comparison_radius} and Table
\ref{tab:dissociation_comparison} completely dissociates by the first rebound
of the simulation.
The dissociation energy is lower than the ionization potentials for any of the
gases presented here.
This allows dissociation to occur sooner than ionization, which may explain
some of the divergence between the effects of ionization and dissociation.
Even though the energy loss is smaller than ionization, deuterium is modeled
as having only one ionization level, whether diatomic or monatomic.
Therefore, the addition of dissociation is similar to adding an ionization
level to deuterium that requires less energy its true ionization level.
By dissociating the deuterium, the net amount of energy that can be lost to
ionization is 1.76 times (based on the ionization potentials in Table
\ref{tab:ionization_potential}) what it would be if dissociation could not
occur, due to the increased number of particles that can be ionized.

In the non-dissociating model, the entirety of the diatomic deuterium ionizes.
In the dissociating model, only 55\% of the particles are ionized, which
results in nearly the same number of ionizations in the bubble due to the
doubling of the number of particles in the bubble.
Even though energy is lost to dissociation in the dissociating model, the net
amount of energy lost to ionization in the dissociating model is 96.7\% that
of the non-dissociating model.
So, for gases with few ionization levels, dissociation does not have a trade
off effect with energy lost to dissociation.

\subsection{Mass Segregation}
\label{subsec:mass_segregation}

\begin{figure*}[t]
    \begin{subfigure}[b]{0.48\textwidth}
        \centering
        \includegraphics[width=\linewidth]{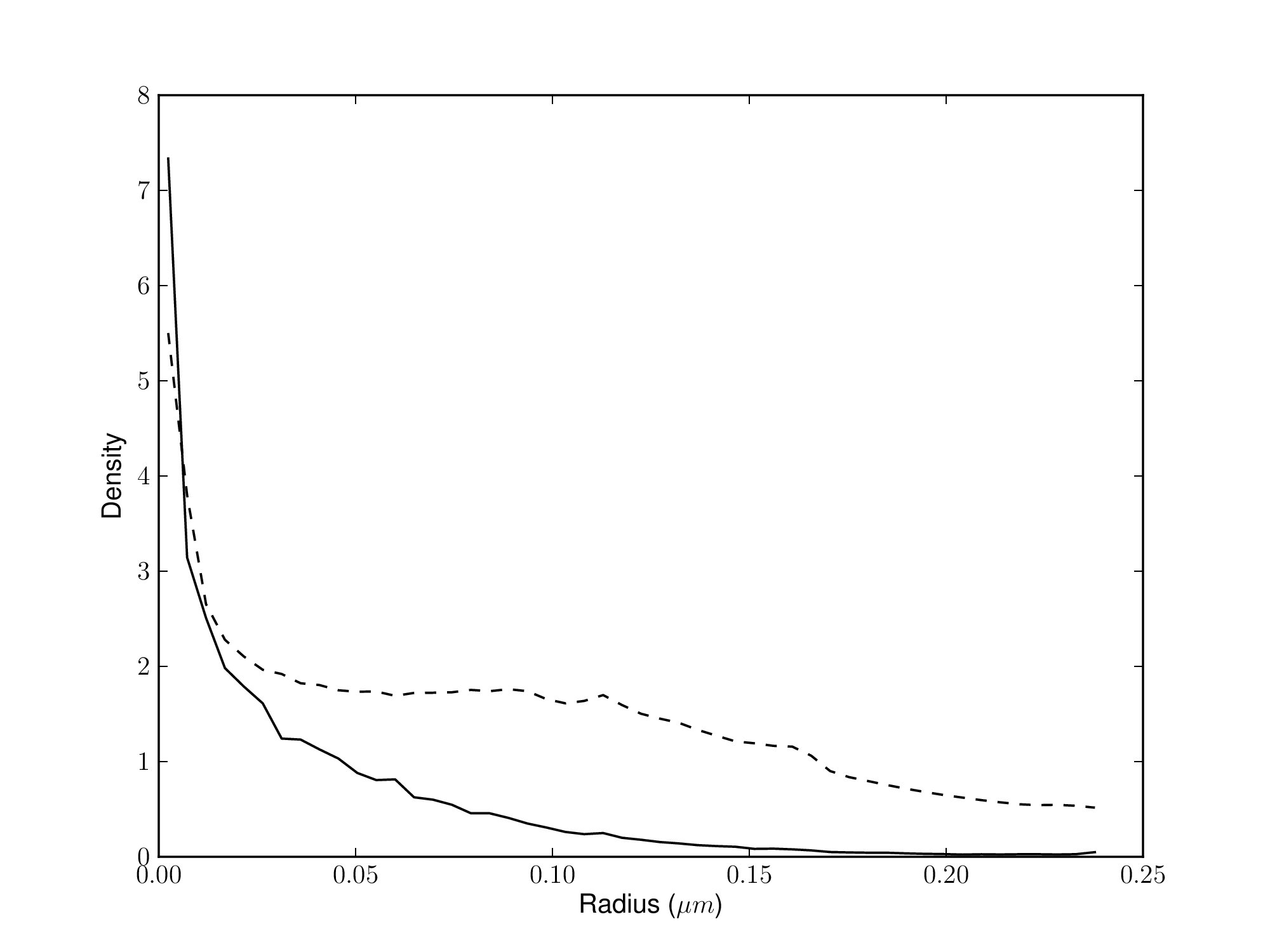}
        \caption{Keller-Miksis}
        \label{fig:density_km}
    \end{subfigure}%
    ~
    \begin{subfigure}[b]{0.48\textwidth}
        \centering
        \includegraphics[width=\linewidth]{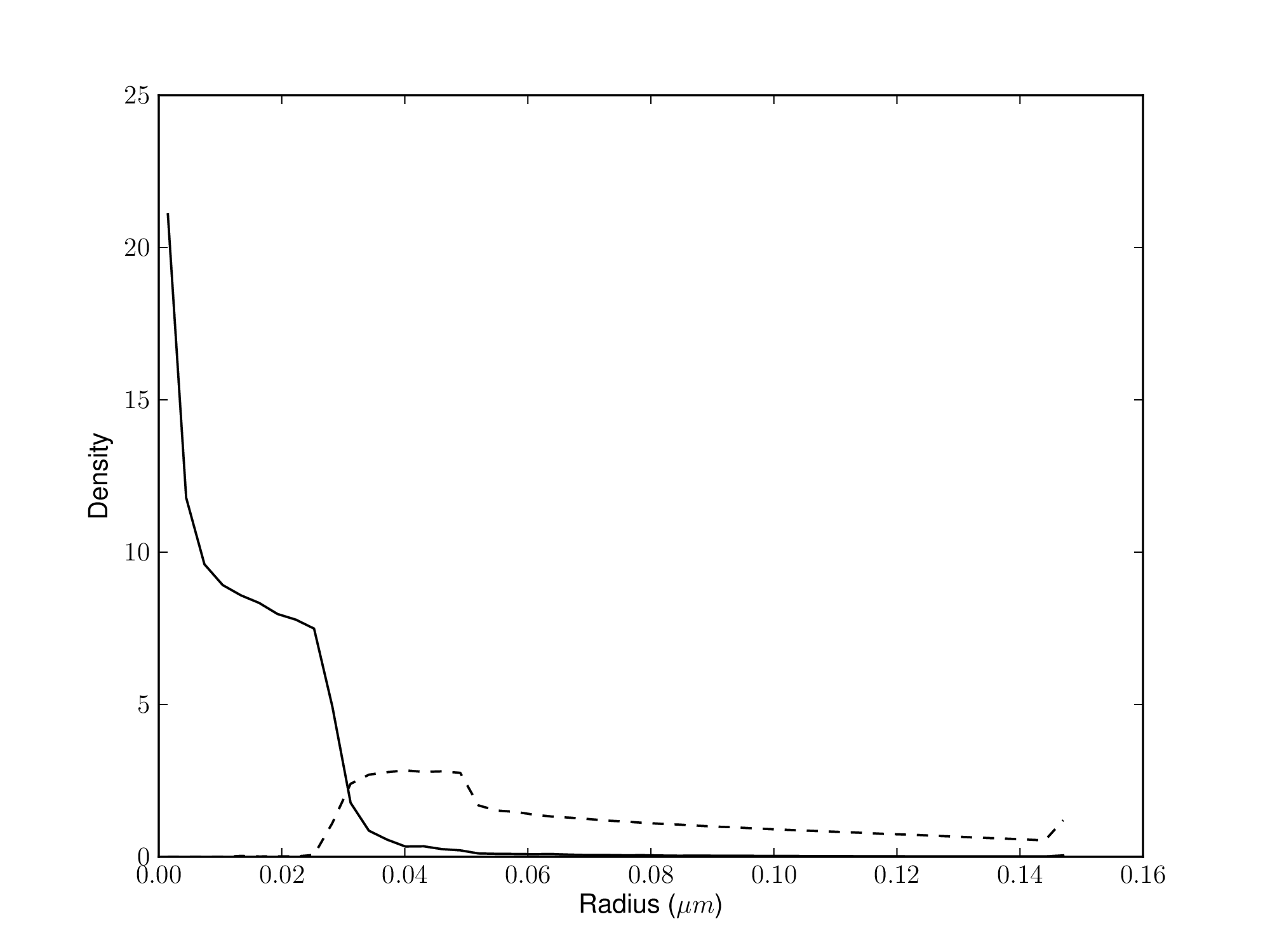}
        \caption{Rayleigh-Plesset}
        \label{fig:density_rp}
    \end{subfigure}
    \caption{Radial density distribution of two simulated bubbles.
        The conditions for these simulations were chosen to match those
        presented in \citet{bass_molecular_2008} using their bubble wall
        equation (Rayleigh-Plesset) and the Keller-Miksis formulation.
        Presented are the densities of the gas densities of the bubbles when
        the shock wave converges at the center of the bubble. The gases are
        xenon (dashed line) and helium (solid line).
        The density values are in units of the mean density of the bubble.}
    \label{fig:rp_comparison}
\end{figure*}

As noted in \citet{bass_molecular_2008}, bubbles containing multiple gases with
sufficiently different masses will gradually segregate during collapse.
This can result in the center of the collapsing bubble being composed of the
light gas almost exclusively, while the heavier gas forms a shell around this
core of light gas.
Mass segregation enhances energy focusing within the bubble and allows the
energy of the collapse to be concentrated on the lightest gas within the system.
Due to how beneficial mass segregation is to energy focusing, it is important to
ensure that mass segregation still occurs in this more complete model.

There are two important factors that influence mass segregation within the
bubble, the difference in mass between the two gas species and the velocity of
the wall.
For the purposes of this paper, we attempt to replicate the results of
\citet{bass_molecular_2008}.
To do so, we collapse an bubble with a 2.2$\mu$m ambient radius using a
1.6 atmosphere sinusoidal driver with a frequency of 30kHz.
Due to the size of the bubble, a 1\degree half-vertex angle was used for the
conical symmetry reduction, which results in a simulation containing
$\approx 136000$ particles with thermal wall conditions.
The gas composition is 90\% xenon and 10\% helium by molar concentration.

Even though the conditions of the simulation here match those presented by
\citet{bass_molecular_2008} mass segregation does not occur in our model, based
on the radial density plots in Figure \ref{fig:rp_comparison}.
This divergence can be attributed to the reduced wall velocity observed in our
model, which results from using a more accurate formulation of the
Rayleigh-Plesset equation, which factors in the effects of surface
tension and damping due to liquid viscosity and acoustic radiation.
The absence of these damping terms has the effect of drastically increasing
the wall velocity.

The simulation using the Rayleigh-Plesset equation from
\citet{bass_molecular_2008} presents a peak wall velocity of 20800 m/s
($\approx$ Mach 16 in water), while the Keller-Miksis formulation used here
has a peak wall velocity of only 3173 m/s.
The Keller-Miksis formulation still presents an increased concentration of He
near the center of the bubble.
The concentration of helium exceeds that of xenon near the center of the
bubble in the KM simulation, but the simulation does not achieve the near
total exclusion of xenon from the center of the bubble that is observed when
using the Rayleigh-Plesset equation.

\citet{storey_mixture_1999} have shown that heat transfer at the gas liquid
interface can also induce mass segregation by creating a temperature gradient
within the bubble.
The thermalizing wall conditions model heat transfer between the gas and liquid
and result in a strong temperature gradient within the bubble
(see Figure \ref{fig:error_temperature_profile}).
The simulations presented here used the thermal wall model to be in keeping with
the results presented in \citet{bass_molecular_2008}.
Even with a strong radial temperature gradient, we were unable to reproduce
the extreme mass segregation of previous works for the lower bubble wall
velocities in our simulation.

\section{Conclusion}
\label{sec:conclusion}

Molecular dynamics methods provide a powerful tool for investigating the
behavior of collapsing bubbles; they provide a way to model the internal gas
system while making fewer assumptions about the behavior of the gas at the
extreme conditions experienced within the bubble.
With the continued increase in processing power of modern computers, more
physically accurate MD models become possible.

In this paper, we have shown how various modeling approaches influence the
evolution of the bubble along with how features of the gas system interplay
with bubble wall motion.
We have shown that gas composition can have a strong influence on the
evolution of the bubble, an effect which is overlooked by using simplified
equations of state or fluid dynamics models.
In addition, we have shown how the choice of molecular model can influence the
motion of the bubble wall.
The effects of wall conditions were considered within our model to examine how
they interplay with other features of the simulation.
Finally, we investigated the effects of dissociation of diatomic gases within
the simulation

All of these features have a quantifiable effect on the motion of the bubble
wall as well as the gas system.
With the simulation pressure now being derived from the gas system, there is
now a possible avenue of verification for these features.
This was mentioned in \citet{kim_validation_2008} as a way to determine the
accommodation coefficient for the wall conditions.
It also necessitates further improvements to the modeling of the gas system,
as the coupling of the gas system and the bubble wall results in certain
modeling choices (e.g.\ choice of molecular cross section model) having a more
significant effect than is apparent in models lacking this coupling.

These improvements show several opportunities for future research.
In particular, a more accurate model of gas behavior at the bubble wall must
be incorporated to achieve results that agree with physical and other
theoretic results.
This could be achieved through the inclusion of an accommodation coefficient
in the same manner as \citet{kim_validation_2008}, which suggests a value of
$\alpha=0.03$ gives strong agreement with theoretical models of nano-sized
argon bubbles.
The appropriate value would have to be determined experimentally in our model
and would likely differ due to the inclusion of several phenomena that are
absent in their model.

Due to the influence that the molecular model has on bubble evolution, more
accurate molecular model for the particles in the system may be necessary.
The values for the VSS model taken from \citet{koura_variable_1992} are only
valid in the 300-2000K range.
The conditions present in our simulations extend well outside that range of
applicability.
In addition, the VSS model does not have an analytical solution for
attractive-repulsive potentials such as the Lennard-Jones potential, as noted
by \citet{hassan_generalized_1993}.
As such, a more robust method such as the Generalized Soft Sphere (GSS) model,
proposed by \citet{fan_generalized_2002}, will provide more physically
accurate results.

An improved ionization model appears to be necessary as well, to accurately
model the bubble after the initial collapse.
To do so, recombination must also be modeled, as the bubble expands and cools.
This cooling would allow electrons to rebind to the particle that lost them,
which would alter the energy of the system.
Though the loss of energy due to ionization happens mostly in the initial
collapse of the bubble, the energy removed from the system has a long term
affect on the trajectory of the bubble wall in our model, as the energy that
went into ionization is permanently lost.
This makes it difficult to assess whether some of the energy lost is due to
ionization or through the bubble wall, as mentioned previously.
This adds further complications when attempting to determine the appropriate
accommodation coefficient for the bubble wall.

The simulation presented here is capable of displaying both strong shock waves
within the gas as well as mass segregation (see Figure
\ref{fig:error_density_profile}).
These phenomena can have a strong influence on energy focusing, but, at least
in the case of shock waves, can also influence the motion of the bubble wall
to some degree.
How these strong heterogeneities interplay with other aspects of the
simulation have not been explored.
In particular, how the presence of a high density region adjacent to the
bubble wall affects heat transfer between the gas and the bubble merits
further exploration.

Finally, regardless of the physical accuracy of the molecular model chosen,
modeling the gas particles as spheres that interact only through collisions is
clearly inaccurate, particularly when the particles ionize.
In such a case, long range electrostatic interactions become significant, and
what effects this will have on the evolution of the bubble is not known.
The inclusion of long range forces is computationally expensive compared to
the algorithm presented here, and it may not be possible to simulate a full
bubble (or even part of one) using such methods.
A fully complete description of the gas would also need to factor in
rotational and vibrational degrees of freedom as well.

\bibliography{paper}{}
\end{document}